\documentclass[notitlepage,prd,twocolumn,showpacs,preprintnumbers,floatfix,nofootinbib,superscriptaddress]{revtex4-2}
\usepackage{amssymb,amsmath,placeins,epsfig,array}
\usepackage[usenames,dvipsnames]{color}
\usepackage[normalem]{ulem}
\usepackage[breaklinks,colorlinks,urlcolor=blue,citecolor=blue,linkcolor=blue]{hyperref}
\usepackage{lineno}
\usepackage{float}
\usepackage{mathptmx}
\usepackage{xcolor, soul}
\sethlcolor{yellow}

\newcommand\ORNL{Oak Ridge National Laboratory, Oak Ridge, Tennessee 37831}
\newcommand\JLab{Thomas Jefferson National Accelerator Facility, Newport News, VA 23606}

\begin{document}

\title{Concept of Very-Asymmetric Lepton Collider for Dark Matter Search}
\author{V.\,S.~Morozov} \thanks{Corresponding author: morozovvs@ornl.gov} \affiliation{\ORNL}
\author{B.~Wojtsekhowski} \affiliation{\JLab}

%
\begin{abstract}
Accelerator-based searches for dark matter are aiming for high sensitivity and need an experimental setup with high luminosity. 
This field of research is often called intensity frontier physics.
One of the best motivated portals of interaction between dark matter and ordinary matter is a dark photon which could be 
observed as a resonance in the invariant mass of the decay products. 
Electron-positron collisions are known to be the cleanest interaction for such a study.
In this paper we propose a scheme for a collider which allows for a luminosity of a few orders of magnitude higher than could be obtained in a conventional symmetric collider and with a few times higher accessible mass than is possible using a positron beam and a fixed target approach.
The key concept is based on asymmetric energies: a high-energy circulating positron beam and a low-energy
high-intensity electron beam, and optimization of the beam interaction region. 
We present here a configuration of a collision region of 10 MeV electron and 4 GeV positron beams.
\end{abstract}
\maketitle

\section{Introduction}

Well motivated searches for a dark photon are aimed at the mass range below 1 GeV~\cite{Arkani-Hamed, BEST, Pospelov, Toro} and require an electron-positron collider with a luminosity level of $10^{34}$~cm$^{-2}$s$^{-1}$. 
A luminosity of this level has, for example, been demonstrated in the collision of 7~GeV electron and 4~GeV positron beams at SuperKeK-B~\cite{Abe2013, Ohnishi2013, KEK}. The invariant energy available for particle production in a collider is called its Center-Of-Momentum (COM) energy $E_{_{COM}}$. The COM energy of a system of two particles with energies, momenta, and masses of $(E_1, \vec{p}_1, m_1)$ and $(E_2, \vec{p}_2, m_2)$, respectively, is given by
\begin{equation}
E_{_{COM}} = \sqrt{(m_1 c^2)^2+(m_2 c^2)^2 + 2(E_1 E_2 -\vec{p}_1\cdot\vec{p}_2)} ,
\label{eq:ecom}
\end{equation}
where $c$ is the speed of light. For a relativistic head-on collider, Eq.~(\ref{eq:ecom}) reduces to $E_{_{COM}} \simeq 2\sqrt{E_1 E_2}$. One of the challenges of low-COM-energy collider studies is that typical collider luminosity drops rapidly with the reduction of energy ($\sim E^2_{_{COM}}$) due to the beam-beam interaction limit and increase in the damping time in the case of stored beams~\cite{Seeman2001}.

We present a concept of a Very Asymmetric linac-ring electron-positron Collider (VAC) that can provide a luminosity of the $10^{34}$~cm$^{-2}$s$^{-1}$ level at a COM energy as low as about 0.1~GeV (see an earlier concept in Ref.~\cite{BW_2017}).
It will allow us to effectively boost the positron energy by a factor of 20 relative to the fixed target configuration considered in Refs.~\cite{BW_2009, BW_2018}. 
It overcomes the low-energy challenges by combining the benefits of a GeV-level flat positron beam in a storage ring with those of a low-energy electron beam from a linac. 
A storage ring provides a high-current naturally-flat small-emittance positron beam that is sufficiently rigid to the beam-beam kicks by the electron beam and has adequate damping rates. 
The low-energy electron beam has an acceptably small power consumption of 1~MW without having to resort to an energy recovery linac. 

The beam quality and match to the flat positron beam result from transporting a round magnetized electron beam and then converting it to a flat beam~\cite{Burov2000, Brinkmann2001, Burov2002, Kim2003} at the interaction point. 
We present and discuss the parameters of such a collider complex based on the demonstrated technologies. 
The effect of the beam-beam interaction on the luminosity is simulated using both weak-strong and strong-strong models~\cite{Bassetti1980, Ziemann1991, Hirata1993, Hirata1995, Papaphilippou1999, Leunissen2000, Qiang2004, Brandt2006b}.
The stability of the stored positron beam against the kink instability and electron beam jitter are also discussed. 
An interaction region schematic is outlined.

\section{Design Concept}
\label{sec:design}
As motivated above, the mass range of interest for the dark photon search is below~1 GeV and requires a high luminosity on the order of $10^{34}$~cm$^{-2}$s$^{-1}$. 
Such a luminosity has been reached in modern lepton colliders~\cite{KEK, Seeman2001, Wienands2007, Abe2013, Olive2014a} and is planned for the future Electron-Ion Collider~\cite{Accardi2016, Adam2021}. 
However, these colliders have COM energies of several to tens of GeV. 
It is challenging to attain this level of luminosity at the low energies of interest for the dark photon search. 
Below the synchrotron radiation power limit, the luminosity typically falls off in symmetric or nearly-symmetric colliders with the square of the COM energy~\cite{Seeman2001, Olive2014a, Zimmermann2018}. 

The luminosity can be parameterized in several seemingly simple forms~\cite{Seeman2001,Brandt2006a,Olive2014b}. It is usually determined by an optimum among multiple aspects of accelerator physics, detector requirements, and engineering constraints. For the dark photon search in the process $e^+e^- \rightarrow \gamma + A'$, a simple calorimeter-based photon detector is sufficient because of the type of the final state. A partial solid angle coverage is also acceptable.

\begin{table*}[t!]
\caption{Parameters of the proposed VAC and their comparison to those of relevant accelerator projects.}
\label{t:vac_par}
\centering
\begin{tabular}{l|c|c|c|c|c}
\hline \hline
Parameter 	& Unit  & \multicolumn{2}{c|}{VAC} & SuperKEK-B			& FNPL~\cite{Piot2006} \\
		&	& Ring 		& Linac	   & LER~\cite{Ohnishi2013}	& 			\\
\hline
particle		 			& 			   & $e^+$		& $e^-$		& $e^+$	& $e^-$			       \\
energy $E$		  		& GeV		   & 4			& 0.01		& 4		& 0.016			       \\
bunch frequency $f_b$	  	& MHz		   & \multicolumn{2}{c|}{248.5}	& 248.5	& 1300			       \\
beam current $I$	   		& A			   & 3.6		& 0.1		& 3.6			& Pulsed			       \\
bunch population $N$		& $\times 10^{10}$	   		& 9.1		& 0.25		& 9.1		& 0.31			       \\
bunch charge $Q_b$		& nC	   		& 14.6		& 0.4		& 14.6		& 0.5			       \\
rms energy spread $\sigma_\delta$	& $\times 10^{-4}$	   	& 8.1		& 2			& 8.1		& 25			       \\
rms bunch length $\sigma_z$		& mm 		   		& 6	   	& 1			& 6		& 1			       \\
rms geom. h/v emittances $\varepsilon_x/\varepsilon_y$
						& nm/nm		& 0.85 / 0.085& 51 / 2.55& 3.2 / 0.009	& 40 / 0.4		       \\
rms norm. h/v emittances $\epsilon_x^n/\epsilon_y^n$ 	& ${\mu \text m}/{\mu \text m}$
						& 6.7 / 0.67	& 1.0 / 0.05	& 25 / 0.07			& 1.3 / 0.013		       \\
IP Twiss beta functions $\beta_x^*/\beta_y^*$ 	& $\text {mm}/\text {mm}$
						& 6 / 0.6		& 0.1 / 0.02	& 32 / 0.27			& 		       \\
rms h/v IP beam sizes $\sigma_x^*/\sigma_y^*$ 	& ${\mu \text {m}}/{\mu \text {m}}$
						& 2.3 / 0.23	& 2.3 / 0.23	& 10 / 0.048			& 		       \\
rms h/v IP ang. spreads $\sigma_x'^*/\sigma_y'^*$ 	& $\text {mrad}/\text {mrad}$
						& 0.37 / 0.37	& 22.2 / 11.1	& 0.32 / 0.18			& 		       \\
h/v beam-beam tune shifts $\xi_x/\xi_y$ 	& 
						& 0.15 / 0.15	& 			& 0.0028 / 0.088		& 		       \\
h/v disruption parameters $D_x/D_y$ 		& $\times 10^3$
						& 			& 15 / 150		& 					& 		       \\
\hline
Luminosity $L$	  					& cm$^{-2}$s$^{-1}$	& \multicolumn{2}{c|}{$3.0 \times 10^{34}$}		
						& 			& 			       \\
\hline \hline
\end{tabular}
\end{table*}

One way to parameterize the luminosity is in terms of the maximum horizontal $\sigma_x^{max}$ and vertical $\sigma_y^{max}$ rms beam sizes, which are often close to the limit due to the dynamic aperture and magnet aperture sizes:
\begin{equation}
L = \frac{N_+ N_- f_b \sigma_x^{max} \sigma_y^{max}}{4\pi l^2 \varepsilon_x^n \varepsilon_y^n} \beta^2\gamma^2 ,
\label{eq:lumi}
\end{equation}
where $N_+$ and $N_-$ are the numbers of particles in the colliding bunches, $f_b$ is the bunch collision frequency, $l$ is the detector space, $\varepsilon_x^n$ and $\varepsilon_y^n$ are the normalized horizontal and vertical rms emittances, respectively, and $\beta$ and $\gamma$ are the relativistic energy factors. 
Note that Eq.~\ref{eq:lumi} assumes several simplifications including equality of the beam energies, match of the beam sizes, lack of the hour-glass effect~\cite{Seeman2001,Brandt2006a,Olive2014b}, and the thin-lens relation between the interaction point (IP) $\beta_{x/y}^*$ and maximum $\beta_{x/y}^{max}$ of $\beta_{x/y}^*\beta_{x/y}^{max}=l^2$. 
Note also that Eq.~\ref{eq:lumi} is not the only constraint. 
Other important constraints coming from the beam-beam interaction are discussed below.

Once selected, the parameters in Eq.~\ref{eq:lumi} (except for $\beta$ and $\gamma$, of course) do not vary significantly with energy. Note that equilibrium normalized emittances $\varepsilon_x^n$ and $\varepsilon_y^n$ drop as the 3rd power of the energy for a lepton beam in a storage ring. 
However, the damping rate drops at the same rate,  becoming impractically long and making the beam sensitive to beam-beam interaction. 
Equation~(\ref{eq:lumi}) illustrates the energy dependence of the luminosity in a symmetric collider. 
This picture is simplified but it is supported by practical experience~\cite{Seeman2001,Olive2014a,Zimmermann2018}.

Our concept for reaching high luminosities at low CM energies is to combine the benefits of a high-energy stored positron beam and a low-energy linac electron beam. 
A sufficiently high current of a positron beam cannot be extracted from a source. 
A storage ring allows one to efficiently accumulate and maintain a high-current positron beam. Synchrotron radiation damping brings the beam to and maintains it in its equilibrium state. 
The ring design and beam parameter choice must be optimized to provide sufficiently small equilibrium emittances and allow for strong focusing of the beam at the IP. 
The positron beam must also be rigid enough and have an adequate damping rate to withstand the beam-beam interaction with the low-energy electron beam. 
Another beneficial feature of a stored positron beam that we are employing is its natural flatness and therefore small transverse 4D emittance. 
This feature has been used in lepton colliders to achieve record luminosities.

An electron source can provide a relatively high current in an electron beam. 
The challenge of providing a high-current electron beam with the small emittances required for a collider is the space charge effect, which makes the beam emittances grow at low energies. 
This problem was first solved in the production of magnetized electron beams for electron cooling applications. 
The generation of a magnetized beam has also been demonstrated using a photo-cathode gun with subsequent acceleration and conversion to a flat beam~\cite{Piot2006}. 
A magnetized beam is produced as a round beam inside a solenoid field. 
This allows one to keep the beam size large, thus mitigating the space charge effect while preserving one of the transverse canonical emittances and therefore keeping the 4D transverse emittance small. 
After acceleration to a high enough energy where the space charge effect is no longer a problem, the magnetized electron beam is converted into a flat beam. 
We do not go into a detailed discussion of the optical design of a round-to-flat beam transformer, since we consider it to be a demonstrated technology.
The beam is adiabatically damped during the acceleration. 
It is then matched to the stored positron beam, which is naturally flat. 
One of the main limitations on the electron beam energy is the power consumption of the electron linac. 

As we show below, the configuration described above allows for a luminosity of about $10^{34}$~cm$^{-2}$s$^{-1}$ using a combination of individually demonstrated beam parameters.


\section{Parameter Choice}
Table~\ref{t:vac_par} lists the VAC parameters we choose using the design strategy outlined in Section~\ref{sec:design}. They are based on the experimentally demonstrated parameters of existing projects with reasonable extrapolations. More specifically, we selected the VAC positron and electron beam parameters to be consistent with those of SuperKEK-B LER and Fermilab/ Northern Illinois Center for Accelerator and Detector Development Photoinjector Laboratory (FNPL), respectively. The SuperKEK-B LER~\cite{Ohnishi2013} and FNPL~\cite{Piot2006} parameters are shown with the VAC parameters in Table~\ref{t:vac_par} for comparison.

The VAC positron beam parameters are nearly identical to those of SuperKEK-B LER. 
The main difference between them is that, for the VAC, we assume about a factor of 4 lower horizontal emittance. 
We expect that this reduction can be achieved with an appropriate low-emittance ring design~\cite{Borland2006,Tsumaki2006}. 
A lower horizontal emittance should, in turn, allow for a lower $\beta_x^*$. 
At the same time, we relax the vertical emittance and $\beta_y^*$ compared to the SuperKEK-B case. 

The VAC electron beam parameters are extrapolated from those of FNPL. Moreover, the electron linac design for a VAC can be directly based on that of FNPL. Its description can be found, for example, in Ref.~\cite{Piot2006}. In the same paper~\cite{Piot2006}, the FNPL group demonstrated both numerically and experimentally that magnetization allows one to mitigate the space charge effect in a beam and reach the invariant emittances of the level needed for a VAC. The FNPL beam was generated in a magnetized state using a photo-cathode gun surrounded by a solenoid. It was then accelerated in the magnetized state and successfully converted into a flat beam as we envision for a VAC. The flat beam at FNPL reached an emittance ratio of about 100, exceeding our assumption of a ratio of 20 for a VAC. Even though FNPL operates in a pulsed mode, it uses the same bunch charge as assumed for a VAC. Therefore, the space charge effects are the same in the two cases. As shown in Table~\ref{t:vac_par}, the FNPL measured single bunch parameters including magnetization, emittances and bunch charge match or even exceed those needed for a VAC. The bunch momentum spread is not critical for the luminosity as long as chromatic aberrations are properly compensated at the IP.  

The choice of the VAC electron beam energy is driven by the physics requirements on the COM energy. The electron energy combined with a reasonable constraint on the beam power of 1~MW sets the limit on the beam current. Reaching a CW electron current of 100~mA needed for a VAC does not present a significant technical challenge. Compared to FNPL, it may only require additional accelerating cavities to provide sufficient power to the beam.

When considering the beam parameters of a collider, it is important to evaluate the impact of mutual beam focusing on the luminosity and beam stability. These questions are discussed in detail in the subsequent sections. The strength of the inter-beam interaction is characterized by the beam-beam tune shifts for stored beams and by the disruption parameters for the linac beams. Their nominal values are provided in Table~\ref{t:vac_par}. The meaning of these parameters is interpreted in the next section.

\section{Beam-Beam Interaction}
\label{sec:bb_interaction}
The collider luminosity and even the stability of a stored colliding beam may be limited by the electro magnetic interaction of the colliding beams at their crossing point. 
To illustrate the scale of this effect, let us consider a flat relativistic bunch characteristic for lepton colliders. 
Assuming that the bunch is Gaussian and its plane is horizontal, it generates a nearly vertical electric field above its central region. 
The peak electric field strength can be estimated as
\begin{equation}
E_y^{max} = \frac{qN}{4\pi\varepsilon_0\sigma_x\sigma_z} ,
\label{eq:maxey}
\end{equation}
where $\varepsilon_0$ is the electric permittivity of free space, $q$ is the particle charge, $N$ is the number of particles in the bunch, and $\sigma_x$ and $\sigma_z$ are the horizontal and longitudinal rms bunch sizes, respectively. 
For example, by applying Eq.~(\ref{eq:maxey}) with the parameters listed in Table~\ref{t:vac_par}, we find that the colliding 10~MeV electrons experience electric fields of the order of $\sim 10$~GV/m. 
The magnetic field generated by the bunch $B_x = \beta E_y /c$ exerts a force on the electrons essentially equal to that of the electric field, since the relativistic parameter $\beta \simeq 1$.

The electric and magnetic fields of a round or elliptical ultra-relativistic bunch can be obtained analytically. 
In the ultra-relativistic limit, the problem is reduced to a two-dimensional Poisson equation and the fields lie in the plane transverse to the bunch direction. They are given by the Bassetti-Erskine formula~\cite{Bassetti1980} in terms of the complex error function. 

The strength of beam-beam interaction experienced by a stored beam is described by the linear beam-beam tune shift parameters
\begin{equation}
\xi_{x,y+} = \frac{r_+ N_- \beta_{x,y+}^*}{2\pi\gamma_+\sigma_{x,y-}^*(\sigma_{x-}^*+\sigma_{y-}^*)} ,
\label{eq:linbb}
\end{equation}
where the indices $+$ and $-$ refer to the beams subjected to and imposing on the beam-beam force, respectively, $r_+$ is the particle classical radius, $N_-$ is the number of particles in the bunch, $\beta_{x,y+}^*$ are the horizontal ($x$) and vertical ($y$) Twiss $\beta$ functions at the IP, and $\sigma_{x,y-}^*$ are the $x$ and $y$ rms beam sizes at the IP. 
The beam-beam parameter has a physical meaning of the betatron tune shift induced by the linear part of the beam-beam force for small-amplitude particles. The $\xi$ parameter describes a linear effect. 
However, the beam-beam force is nonlinear in nature and causes a betatron tune spread in the beam. 
The $\xi$ parameter is used as a scale of the beam-beam induced tune spread. 
Only a limited tune spread can be accommodated in the storage ring's tune stability region. 
The details of the tune spread and what the ring can tolerate depend on the particular ring design. 
However, the practical experience has been that $\xi$ is limited to a maximum of about $0.15$ for lepton rings with strong synchrotron damping. 
This is the value we adopt for the stored positron beam in Table~\ref{t:vac_par}.

Note that, while the nominal $\xi$ in Table~\ref{t:vac_par} is somewhat high, it does not accurately describe the strength of the electron impact on positrons. Equation~(\ref{eq:linbb}) assumes that both bunches are rigid. However, as shown by the simulations described later in this paper, the electrons oscillate quickly inside the positron beam, which significantly averages out the transverse kick they impose on the positrons. Moreover, the electron beam is quickly blown up by the positron beam. This results in an effective transverse kick experienced by the positrons from the electrons that is much lower than expected from the nominal $\xi$ value obtained by a na\"ive application of Eq.~(\ref{eq:linbb}).

The beam-beam constraints are different for the linac beams. 
Since there is no need to maintain the particle motion stability over multiple turns, linac beams allow for much stronger beam-beam effects. 
The strength of the beam-beam interaction in this case is characterized by the disruption parameters
\begin{equation}
D_{x,y-} = \frac{\sigma_{z+}}{f_{x,y-}} ,
\label{eq:disrupparfoc}
\end{equation}
where $f_{x,y-}$ are the horizontal and vertical focal lengths of the linac beam in the fields of the opposing beam. 
The focal lengths are given by
\begin{equation}
f_{x,y-} = \frac{\beta_{x,y-}^*}{4\pi \xi_{x,y-}}. 
\label{eq:foclen}
\end{equation}
Combining Eqs.~(\ref{eq:linbb}), (\ref{eq:disrupparfoc}), and (\ref{eq:foclen}), the disruption factors can be expressed as
\begin{equation}
D_{x,y-} = \frac{2 r_- N_+ \sigma_{z+}}{\gamma_-\sigma_{x,y+}^*(\sigma_{x+}^*+\sigma_{y+}^*)} .
\label{eq:disrupparsigma}
\end{equation}

The physical interpretation of a small-valued disruption parameter $D_{x,y-} \ll 1$ is the relative change in the incoming particle's transverse position inside the length of the opposing bunch. 
In the situation when $D_{x,y-} \ll 1$, the opposing bunch acts as a thin lens. 
This notion breaks down when  $D_{x,y-} \gg 1$. 
The incident particle then oscillates inside the plasma of the opposing bunch. 
The number of these oscillations can be estimated as~\cite{Hao2010}.
\begin{equation}
n_{x,y-}^{osc}= \frac{\sqrt{D_{x,y-} }}{(2\pi)^{3/4}} .
\label{eq:oscnumber}
\end{equation}

The electron beam disruption parameters for the parameters of our proposed collider scheme are given in Table~\ref{t:vac_par}. Application of Eq.~(\ref{eq:oscnumber}) gives that they correspond to about 31 and 97 oscillations in the horizontal and vertical directions, respectively. 
The horizontal and vertical focal lengths obtained using Eq.~(\ref{eq:foclen}) are 0.22~$\mu$m and 22~nm, respectively. 
The luminosity expression in Eq.~(\ref{eq:lumi}) assumes that beam-beam interaction is weak enough that the change in the particle transverse positions during collision can be ignored. 
This assumption is valid for the positron beam whose particles nominally undergo about one tenth of an oscillation. 
Technically, even the tune shift interpretation of Eq.~(\ref{eq:linbb}) is only valid under the assumption of negligible transverse particle shift, but its magnitude still indicates weakness of the beam-beam effect on the positron beam. 
Clearly, the electron beam dynamics during collision and the resulting luminosity must be simulated numerically. 

\section{Simulation}

We first use a weak-strong (WS) model to simulate the electron-positron collision~\cite{Bassetti1980, Ziemann1991, Hirata1993, Hirata1995, Papaphilippou1999, Leunissen2000, Qiang2004, Brandt2006b}. 
Following the above reasoning, we assume that the positron bunch is frozen. 
The particle density distribution of the electron bunch is modeled using a sufficiently large number of the order of $10^4$ of randomly-generated representative particles. We assume that the frozen positron and initial electron distributions are both Gaussian in all dimensions with rms widths specified in Table~\ref{t:vac_par}. 

The electric and magnetic fields generated by the positron bunch are calculated using the Bassetti-Erskine equations~\cite{Bassetti1980}. 
We simulate the electron dynamics in these fields using synchro-beam mapping~\cite{Hirata1993}. 
Below, we briefly summarize the main steps of this algorithm. 
The positron bunch is split into a number of longitudinal slices. 
Each of the representative electrons sequentially interacts with all of the slices in the order they are encountered. 
The longitudinal position of each of the electron-slice interactions is calculated using the time coordinates of the electron $t_{i-}$ and of the slice $t_{j+}$ in their respective bunches as
\begin{equation}
s_{ij} = -c(t_{i-} - t_{j+})/2 .
\label{eq:longpos}
\end{equation}
The electron coordinates are propagated from the IP to the interaction point as
\begin{equation}
X_{ij-} = x_{i-}+x_{i-}' s_{ij},~~~Y_{ij-} = y_{i-}+y_{i-}' s_{ij},
\label{eq:elcoord}
\end{equation}
where $x_{i-}/y_{i-}$ and $x_{i-}'/y_{i-}'$ are the electron horizontal/vertical coordinates and angles at the IP where $s=0$. The positron slice rms widths are propagated to the same point as
\begin{eqnarray}
\Sigma_{xij+} & = & \sqrt{\sigma_{x+}^2+\sigma_{x'+}^2 s_{ij}^2}, \nonumber \\
\Sigma_{yij+} & = & \sqrt{\sigma_{y+}^2+\sigma_{y'+}^2 s_{ij}^2},
\label{eq:poswidths}
\end{eqnarray}
where the lower-case $\sigma$ symbols refer to the parameter values at the IP.

Interaction is modeled as a thin-lens kick imposed on the electron by the electric and magnetic fields of the positron slice. The electric field $\vec{E}_{ij+}^{BE}$ at the propagated position $(X_{ij-},\,Y_{ij-},\,s_{ij})$ of the electron is calculated using the Bassetti-Erskine equations with the propagated positron slice parameters $\Sigma_{xij+}$ and $\Sigma_{yij+}$. The slice population $n_{j+}$ is obtained based on the time interval of the slice $[-t_{j+}^{lb},\: -t_{j+}^{ub}]$ where $-t_{j+}^{lb}$ and $-t_{j+}^{ub}$ are the lower and upper bounds of the time interval, respectively. It is convenient to convert time to length units as $[z_{j+}^{lb},\: z_{j+}^{ub}] \equiv -c [t_{j+}^{lb},\: t_{j+}^{ub}]$. Using a longitudinal positron density distribution $\rho_+(z) = \int \rho_+(x,y,z) dxdy$, the number of positrons in the $j$th slice $n_{j+}$ is given by
\begin{equation}
n_{j+} = N_+ \int_{z_{j+}^{lb}}^{z_{j+}^{ub}} \rho_+(z) dz . 
\end{equation}
Assuming that $\rho_+(z)$ is Gaussian, $n_{j+}$ can be calculated using the error function as
\begin{equation}
n_{j+} = N_+ \frac{1}{2}\left[\textrm{erf} \left( \frac{z_{j+}^{ub}}{\sqrt{2}\sigma_{z+}} \right) - \textrm{erf}\left( \frac{z_{j+}^{ub}}{\sqrt{2}\sigma_{z+}} \right) \right].
\label{eq:slicepop}
\end{equation}

The electron momentum change resulting from its interaction with the slice is calculated using the Lorentz force equation as
\begin{eqnarray}
\Delta \vec{p}_{ij-} & = & \vec{F}_{ij} \Delta  t_{j+} \nonumber \\
& = & \frac{1}{2c} \vec{F}_{ij} \Delta  z_{j+}  \nonumber \\
& = & -\frac{e}{2c} (\vec{E}_{ij+}^{BE} + c \vec{\beta}\times \vec{B}_{ij+}^{BE}) \Delta  z_{j+} \nonumber \\
& \simeq & -\frac{e}{c} \vec{E}_{ij+}^{BE} \Delta  z_{j+} \nonumber \\
& = & -\frac{e}{c} \frac{\vec{E}_{ij+}^{BE}}{N_+ \rho_{zj+}} n_{j+} .
\label{eq:momkick}
\end{eqnarray}
The quantity $\vec{E}_{ij+}^{BE}/(N_+ \rho_{zj+})$ in Eq.~(\ref{eq:momkick}) is independent of $N_+$ and $\rho_{zj+}$. That dependence is folded into $n_{j+}$. 

The electron transverse angular coordinates are updated as 
\begin{equation}
\vec{\tilde{x}}_{i-}' = \vec{x}_{i-}' + \Delta \vec{p}_{ij-}/p_{zi-} .
\end{equation}
The longitudinal momentum component $p_{zi-}$ is also updated according to the algorithm described in~\cite{Hirata1993} to keep the entire process symplectic but this longitudinal effect is small. Finally, the transverse spacial coordinates are projected back to the IP using the updated angles
\begin{equation}
\vec{\tilde{x}}_{i-} = \vec{X}_{ij-} - \vec{\tilde{x}}_{i-}' s_{ij}.
\end{equation}
The process continues for all representative electrons with each electron sequentially kicked by the positron bunch slices from first to last.

Each electron-slice interaction contributes the following amount to the luminosity expressed using the quantities in Eqs.~(\ref{eq:elcoord}) and (\ref{eq:poswidths}).
\begin{equation}
L_{ij} = \frac{n_{j+} N_- f_b}{2\pi n_- \Sigma_{xij+} \Sigma_{yij+} } 
\exp \left( -\frac{X_{ij-}^2}{2 \Sigma_{xij+}^2} -\frac{Y_{ij-}^2}{2 \Sigma_{yij+}^2} \right) ,
\end{equation}
where $n_-$ is the number of representative electrons used to model the total number of electrons per bunch $N_-$. The total luminosity is then
\begin{equation}
L = \sum_{i=1}^{n_-} \sum_{j=1}^{m_+} L_{ij} ,
\end{equation}
where $m_+$ is the number of positron slices.

Let us next estimate how finely the positron bunch has to be sliced to accurately model the electron dynamics inside of it. 
The criterion that needs to be satisfied to ensure validity of the described algorithm is that the relative change of the electron's transverse position inside the slice is much less than unity. 
Casting this condition in terms of the slice length $\Delta z_{j+} = z_{j+}^{ub} - z_{j+}^{lb}$ and its focal parameters $f_{x,y-}^{j+}$ specified by Eq.~(\ref{eq:foclen}), we can write
\begin{eqnarray}
\Delta z_{j+} & \ll & \min\left\{ f_{x,y-}^{j+} \right\} = f_{y-}^{j+} \nonumber \\
 & = & \min_i \left \{ 
\frac{\gamma_- \Sigma_{yij+}^2 \left( \Sigma_{xij+}^2 + \Sigma_{yij+}^2 \right) }{2 r_- n_{j+}} \right\},
\label{eq:sl_len_lthan_focp}
\end{eqnarray}
since in most cases including ours $f_{y-}^{j+} < f_{x-}^{j+}$. Using the facts
\begin{eqnarray}
\sigma_{x+}^* & \leq & \Sigma_{xij+},~~~\sigma_{y+}^* \leq \Sigma_{yij+}, \\
n_{j+} & \simeq & N_+ \rho_+(z_{j+}^{lb}) \Delta z_{j+} \leq N_+ \rho_+(0) \Delta z_{j+} \nonumber \\
&& = \frac{N_+ \Delta z_{j+}}{\sqrt{2\pi}\sigma_{z+} }
\end{eqnarray}
in Eq.~(\ref{eq:sl_len_lthan_focp}), it is sufficient to satisfy
\begin{equation}
\Delta z_{j+} \ll \sqrt{\frac{\sqrt{2\pi} \gamma_- \sigma_{y+}^* 
\left( \sigma_{x+}^* + \sigma_{y+}^* \right) \sigma_{z+}}{2 r_- N_+}} .
\label{eq:sl_len_cond}
\end{equation}
Using Eqs.~(\ref{eq:disrupparsigma}) and (\ref{eq:sl_len_cond}), the condition for the number of slices per rms bunch length can be expressed as
\begin{equation}
\frac{\sigma_{z+}}{\Delta z_{j+}} \gg \sqrt{\frac{D_{y-}}{\sqrt{2\pi}}} .
\label{eq:no_sl_cond}
\end{equation}
For the parameters listed in Table~\ref{t:vac_par}, Eq.~(\ref{eq:no_sl_cond}) gives $\sigma_{z+}/\Delta z_{j+} \gg 244$. 

When using equally-sized slices, the limit on the slice size is determined by the highest particle density at the center of the bunch. 
At lower particle densities away from the bunch center, this size is smaller than necessary, leading to a low computation efficiency. 
An unnecessarily large fraction of time is spent on tracking through the bunch tails that have a lower impact on the beam dynamics and luminosity than the central part of the bunch. 
On the other hand, due to the low magnetic rigidity of the electron beam, there may be a non-negligible effect of the tails extending beyond the $\pm 3\sigma_{z+}$ or even $\pm 6\sigma_{z+}$ range. 
To resolve this issue, we use an adaptive slicing approach with variable slice sizes. 
The bunch is symmetrically divided into $m_+$ slices in such a way that all slices contain equal fractions of the bunch population of $1/m_+$. The $z_{j+}^{ub}$ coordinates of the upper bounds of all slices are found by solving
\begin{equation}
\int_{-\infty}^{z_{j+}^{ub}} \rho_+(z) dz = \frac{j}{m_+} ,
\end{equation}
where $j = 1, \ldots , m_+$. Clearly, $z_{m+}^{ub} = +\infty$. Since the segments are adjacent, the lower bounds $z_{j+}^{lb} = z_{(j-1)+}^{ub}$ for $j>1$ and $z_{1+}^{lb} = -\infty$. The effective center of the slice $z_{j+} = -c t_{j+}$ for use in Eq.~(\ref{eq:longpos}) is chosen to be the median point of the particle density along the slice that can be obtained from
\begin{equation}
\int_{-\infty}^{z_{j+}} \rho_+(z) dz = \frac{j-1}{m_+} .
\end{equation}

\begin{figure}[t]
   \centering
   \includegraphics*[width=\columnwidth]{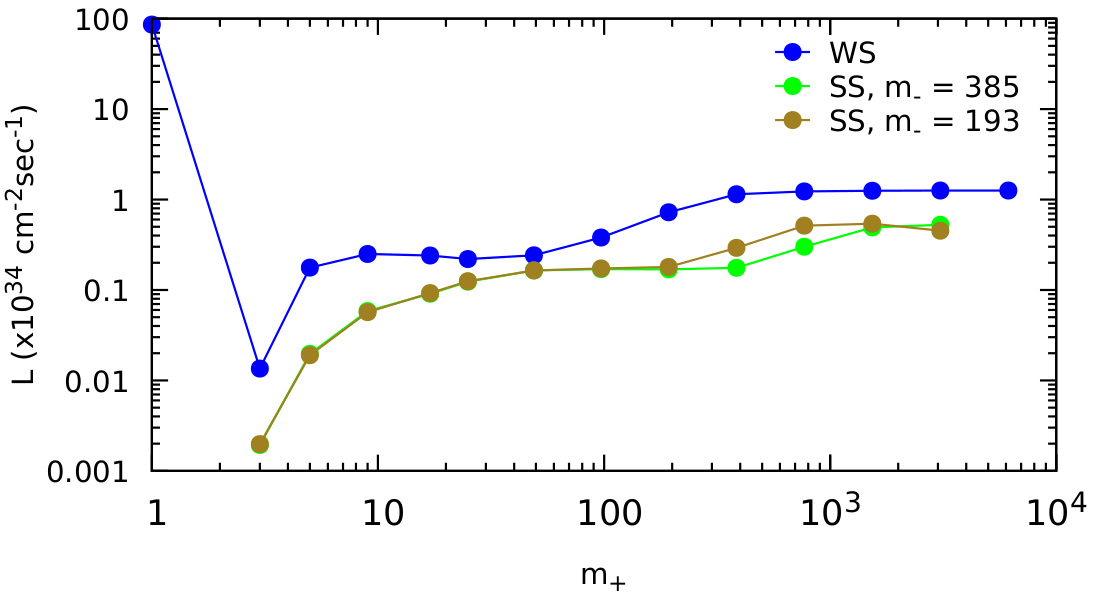}
   \caption{Luminosity convergence with the number of positron bunch slices $m_+$ in a WS case (blue), an SS case with the number of electron slices $m_-=385$ (green), and an SS case with $m_-=193$ (olive).}
   \label{fig:lumi_vs_nslices_0mrad}
\end{figure}

We apply the algorithm described above to calculate the electron dynamics and the resulting electron-positron collision luminosity using the parameters listed in Table~\ref{t:vac_par}. 
We use $n_- = 10^4$ representative electrons. Note that, in the WS case, $n_-$ is not the number of macro-particles. It is the number of particles representing the electron distribution. It does not have any impact on the simulated dynamics and only affects the accuracy of luminosity calculation. Its statistical error is of the order of 1\%.
Their coordinates are randomly generated according to their rms distribution widths in Table~\ref{t:vac_par} with $3\sigma$ cutoff. To ensure that the number of positron slices $m_+$ is sufficiently large to provide an accurate luminosity result, we increase $m_+$ until the luminosity converges to a constant value. 
Figure~\ref{fig:lumi_vs_nslices_0mrad} shows that the luminosity plateaus at about $1.26 \times 10^{34}$~cm$^{-2}$sec$^{-1}$ for $m_+ \gtrsim 10^3$. This $m_+$ limit is consistent with our expectation according to Eq.~(\ref{eq:no_sl_cond}). 
We conservatively choose $m_+ = 6145$ for further WS studies.

The point at $m_+ = 1$ in Fig.~\ref{fig:lumi_vs_nslices_0mrad} corresponds to the case of infinitesimally short bunches when the luminosity can simply be calculated as
\begin{equation}
L = \frac{N_+ N_- f_b}{4 \pi \sigma_x^* \sigma_y^*} .
\label{eq:lumi_short_bunches}
\end{equation}
For the parameters of Table~\ref{t:vac_par}, Eq.~\ref{eq:lumi_short_bunches} gives a luminosity of $L = 8.5 \times 10^{35}$~cm$^{-2}$sec$^{-1}$. Our Monte-Carlo simulation using the algorithm described above finds $L = 8.6 \times 10^{35}$~cm$^{-2}$sec$^{-1}$, as shown in Fig.~\ref{fig:lumi_vs_nslices_0mrad}. This is a good agreement, which is within the systematic error determined by the chosen number of representative electrons. This provides a sanity check of the validity of our calculation. 

Once electron dynamics in the fields of the positron bunch is accurately taken into account by using a sufficiently large number of positron bunch slices, the resulting realistic luminosity estimate is almost two orders of magnitude lower than that of Eq.~(\ref{eq:lumi_short_bunches}) but it still is rather high for this energy range at $L = 1.26 \times 10^{34}$~cm$^{-2}$sec$^{-1}$. This value is adequate for the physics goals described earlier.

After establishing an initial benchmark with a simplified by physically valid WS model, we verify it by strong-strong (SS) beam-beam simulations. Accurate strong-strong beam-beam simulations are challenging in the considered parameter range due to a large number of macro-particles and a high grid density required for accurate modeling of the electron dynamics inside a positron bunch. We use an existing BeamBeam3D code~\cite{Qiang2004} to repeat the WS convergence study. The BeamBeam3D SS simulation results are shown in Fig.~\ref{fig:lumi_vs_nslices_0mrad} for comparison to the WS calculations. Each of the colliding bunches is modeled in BeamBeam3D using about 33.6 million macro-particles on an $x \times y \times z$ grid of $32 \times 32 \times 24576$. These parameters and the maximum slice numbers were limited by the available computer memory but the SS results converge within the achievable parameter settings. Even though the SS calculations give a factor of 2 lower luminosity, given the numerical and model limitations of the two calculations, we consider the WS and SS results to be in a reasonable agreement. 

\begin{figure}[t]
   \centering
   \includegraphics*[width=\columnwidth]{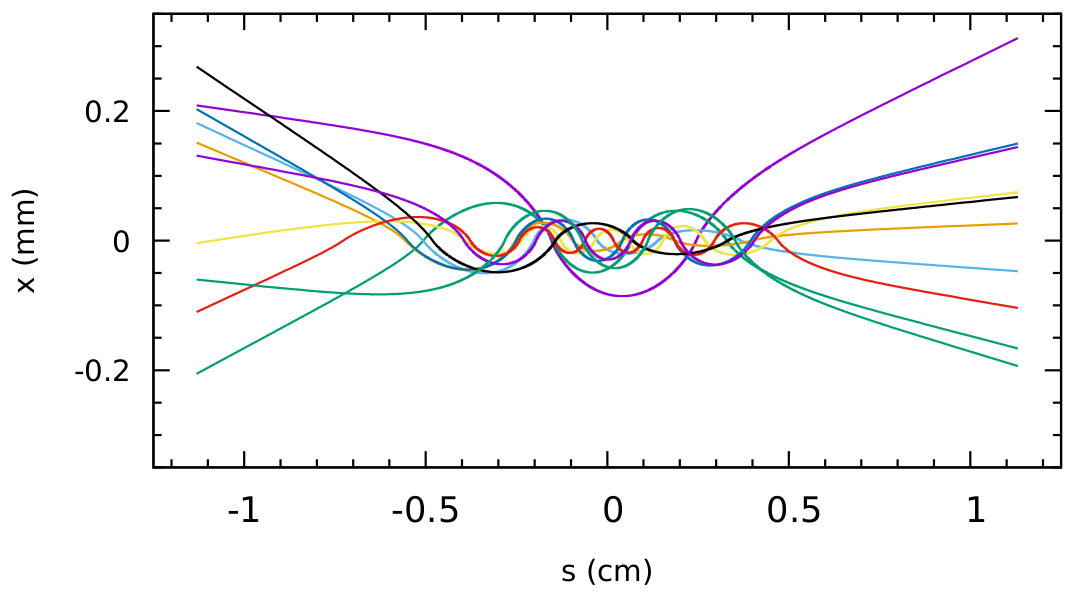}
   \includegraphics*[width=\columnwidth]{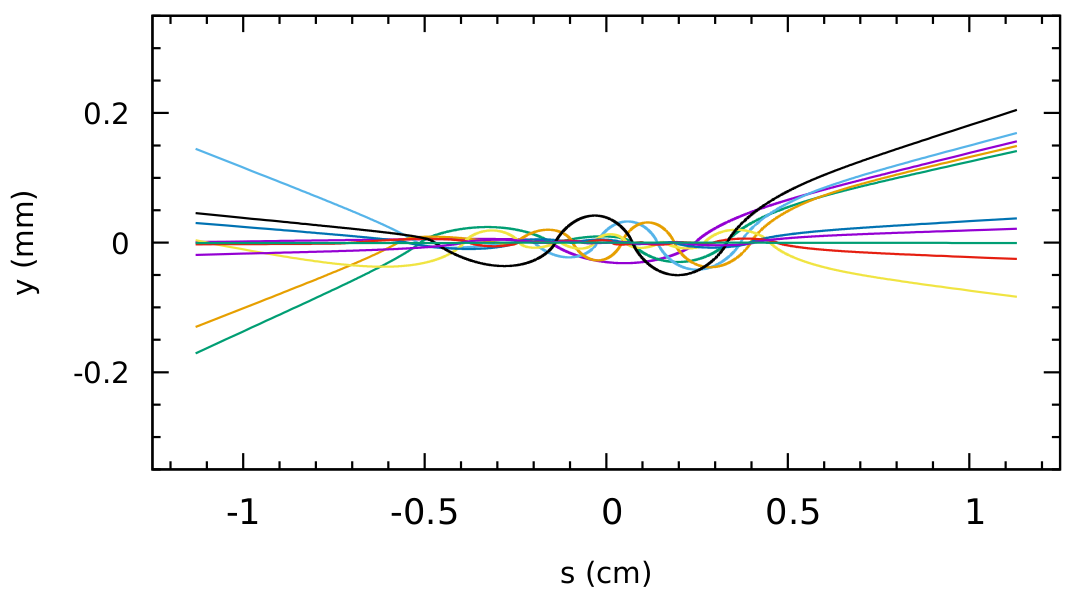}
   \caption{Horizontal (top) and vertical (bottom) trajectories of ten sample electrons inside a positron bunch in the WS case. $s$ is the electron longitudinal position in the lab frame. $s=0$ corresponds to the IP, i.e. the crossing point of the bunch centers. The initial electron coordinates were randomly generated with the distribution parameters specified in Table~\ref{t:vac_par}. The positron bunch parameters are also given in Table~\ref{t:vac_par}. The positron bunch was divided into $m_+ = 6145$ slices.}
   \label{fig:10rand_ele}
\end{figure}

The motion of individual electrons inside a positron bunch in the WS case is illustrated in Fig.~\ref{fig:10rand_ele}. 
It shows the horizontal and vertical trajectory components of ten electrons which were randomly sampled from the distribution used in the WS luminosity calculations of Fig.~\ref{fig:lumi_vs_nslices_0mrad}. 
The trajectories are shown for the maximum number of positron slices $m_+ = 6145$. 
The electrons rapidly converge and diverge before and after interaction with the positron bunch, respectively. 
The electron trajectories rapidly oscillate but remain focused during interaction with the central part of the positron bunch. 

\begin{figure}[t]
   \centering
   \includegraphics*[width=\columnwidth]{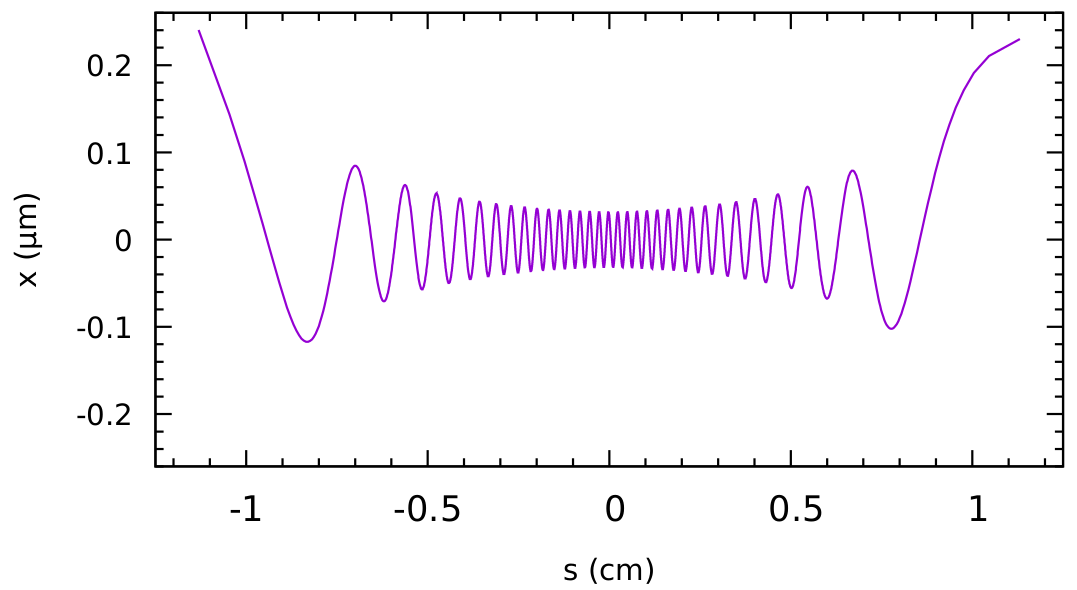}
   \includegraphics*[width=\columnwidth]{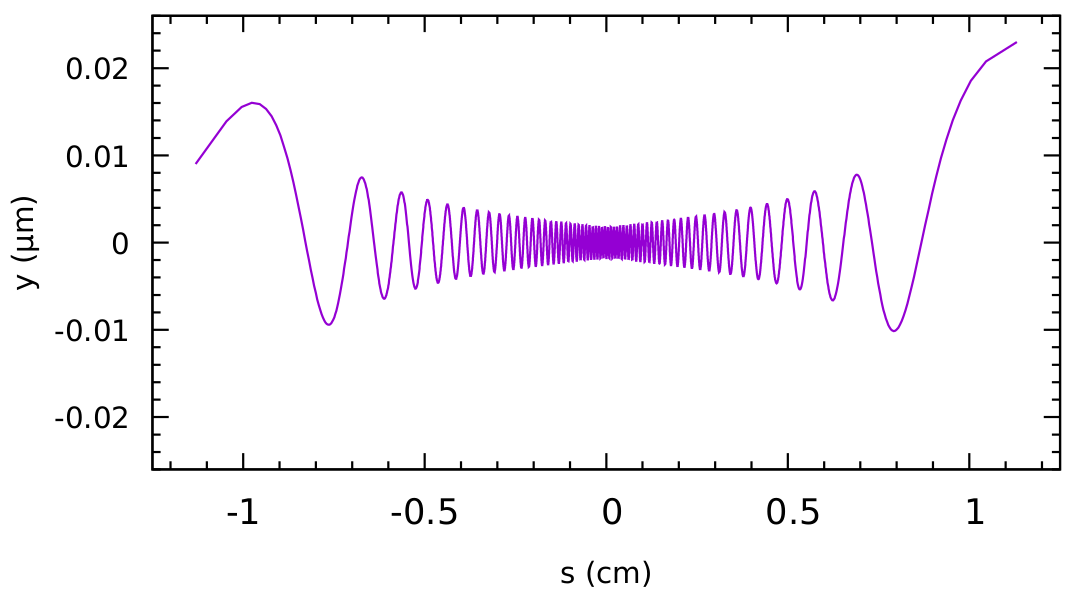}
   \caption{Horizontal (top) and vertical (bottom) trajectories of an electron with coordinates of $x = 0.1\sigma_x^*$, $y = 0.1\sigma_y^*$, and $x^\prime = y^\prime =0$ in the middle of the positron bunch in the WS case. $s$ is the electron longitudinal position in the lab frame. $s=0$ corresponds to the IP. The electron and positron parameters are listed in Table~\ref{t:vac_par}. The number of the positron slices is $m_+ = 6145$. }
   \label{fig:sm_amp_ele}
\end{figure}

\begin{figure}[t]
   \centering
   \includegraphics*[width=\columnwidth]{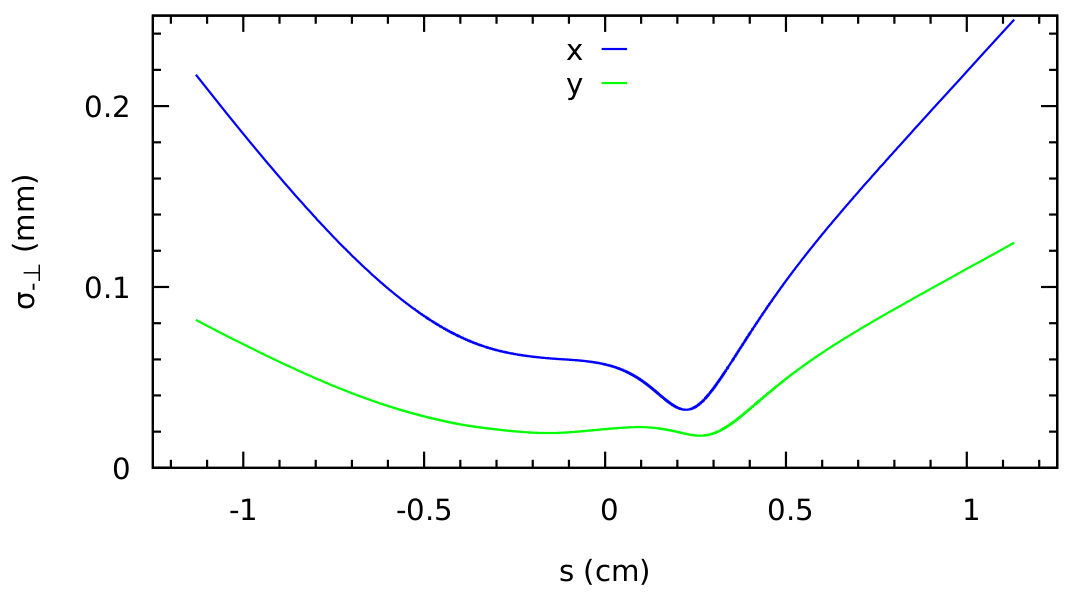}
   \caption{Evolution of the rms horizontal ($x$) and vertical ($y$) sizes ($\sigma_{-\perp}$) of the electron bunch during its propagation through the positron bunch in the WS case. $s$ is the longitudinal position of the electron bunch center in the lab frame. $s=0$ corresponds to the IP. The electron and positron parameters are listed in Table~\ref{t:vac_par}. The number of the positron slices is $m_+ = 6145$. }
   \label{fig:ele_size}
\end{figure}

\begin{figure}[t]
   \centering
   \includegraphics*[width=\columnwidth]{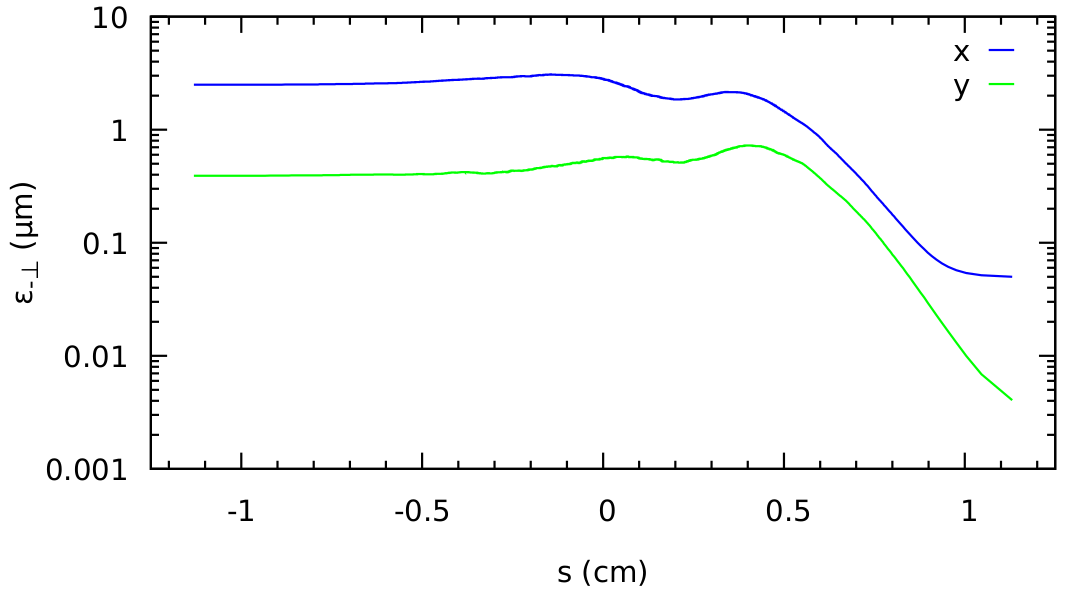}
   \caption{Evolution of the rms horizontal ($x$) and vertical ($y$) geometric emittances ($\sigma_{-\perp}$) of the electron bunch during its propagation through the positron bunch in the WS case. $s$ is the longitudinal position of the electron bunch center in the lab frame. $s=0$ corresponds to the IP. The electron and positron parameters are listed in Table~\ref{t:vac_par}. The number of the positron slices is $m_+ = 6145$. }
   \label{fig:ele_emit}
\end{figure}

It may appear from Fig.~\ref{fig:10rand_ele} that the number of electron oscillations is significantly smaller than what we estimated using Eq.~(\ref{eq:oscnumber}). The reason for this is that Eq.~(\ref{eq:oscnumber}) is only valid for small-amplitude electrons. Figure~\ref{fig:sm_amp_ele} demonstrates this by plotting the horizontal and vertical trajectory components of a small-amplitude electron. The electron trajectory is offset by one tenth of the rms beam size in both $x$ and $y$ and has zero slope in the middle of the positron bunch. The numbers of $x$ and $y$ oscillations in Fig.~\ref{fig:sm_amp_ele} are now consistent with the predictions of Eq.~(\ref{eq:oscnumber}). This is another benchmark of our calculation. Figures~\ref{fig:ele_size} and \ref{fig:ele_emit} show evolutions of the transverse sizes and emittances, respectively, of the electron bunch during its propagation through the positron bunch in the WS case.

\begin{figure}[t]
   \centering
   \includegraphics*[width=\columnwidth]{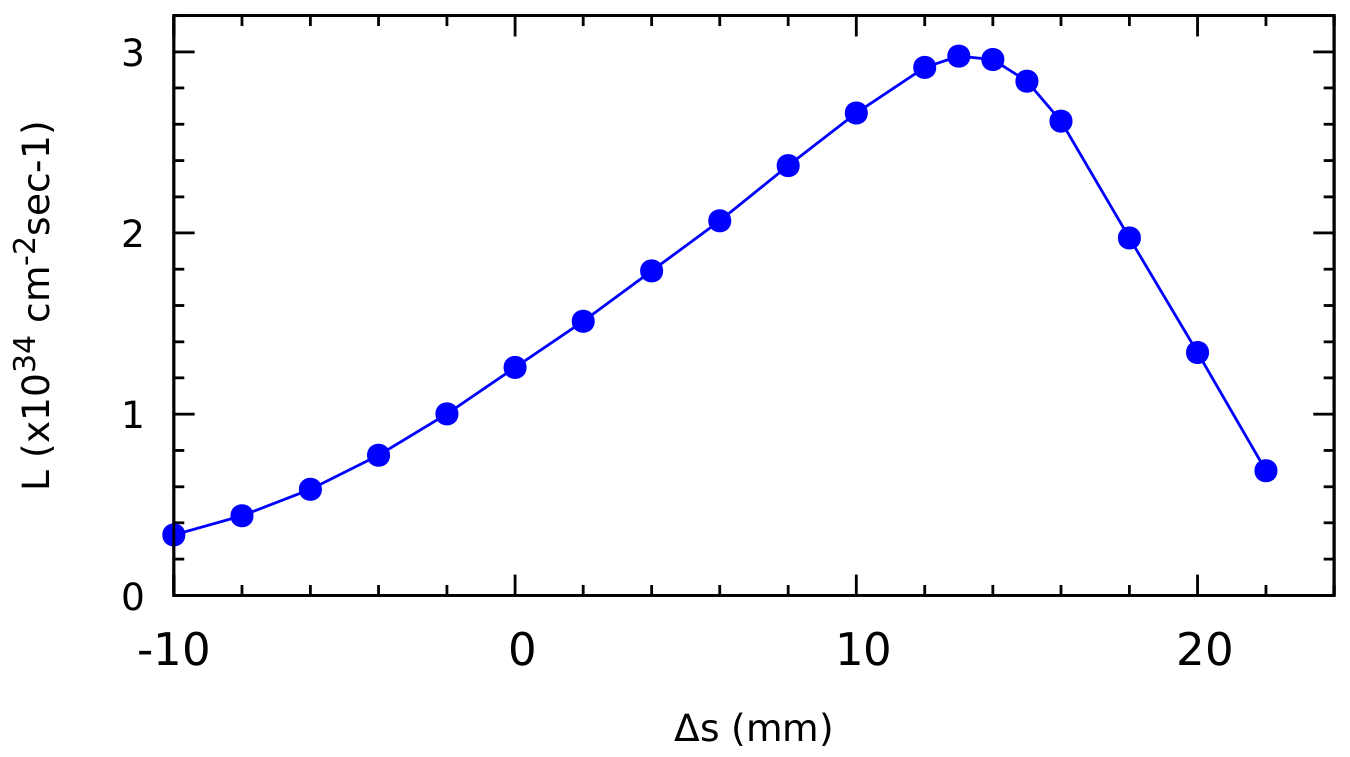}
   \caption{Luminosity as a function of the longitudinal offset of the electron focal point with respect to the positron one in the WS case. A positive $\Delta s$ value corresponds to a shift in the outgoing positron beam direction. The electron and positron parameters are listed in Table~\ref{t:vac_par}. The number of the positron slices is $m_+ = 6145$. }
   \label{fig:lumi_vs_zec}
\end{figure}

We next study the sensitivity of luminosity to collision setup parameters. 
We consider one at a time the impacts of a longitudinal shift of the electron focal point, a transverse offset of the electron beam axis, and its angular misalignment. 
The longitudinal electron bunch shift is set with respect to the positron focal point while the transverse offset and angle of the electron bunch centroid are specified with respect to the electron focal point. 
These parameters are nominally defined for the initial electron bunch prior to its interaction with the positron beam. 
This means that, with the beam-beam interaction taken into account, the coordinates of the electron bunch centroid at the electron focal point are in general not the same as their nominal unperturbed settings. 
We define the parameters this way because that is how they are set in practice, since one does not have access to the actual electron bunch centroid coordinates during bunch interaction.   

Figure~\ref{fig:lumi_vs_zec} shows the luminosity as a function of the longitudinal position of the electron focal point with respect to the positron one in the WS case. 
The point at $s = 0$~m is the same as the $m_+ = 6145$ point in Fig.~\ref{fig:lumi_vs_nslices_0mrad}. 
The $s$ axis in Fig.~\ref{fig:lumi_vs_zec} points along the positron beam direction. 
Moving the electron focal point in the direction opposite to the positron beam lowers the luminosity, while its shift along the positron beam first increases the luminosity up to a maximum and then makes it fall off again. 
This behavior of luminosity can be explained by the change in the optical matching of the electron beam to the positron bunch. 
The positron bunch serves as a strong focusing element for electrons. It focuses them in both dimensions. 
At the linear level, there exists an optics solution symmetric about the positron bunch center. 
This solution corresponds to the tightest focusing of the electron beam inside the positron bunch when the electrons are held as close together as possible throughout the interaction process. 

Moving the electron focal point back and forth changes how well the electron distribution is matched to the symmetric solution discussed above. 
The highest luminosity point in Fig.~\ref{fig:lumi_vs_zec} corresponds to the closest match of electrons to the optimal solution, while shifting the position from this point in either direction makes this match worse, thus increasing the electron beam size inside the positron bunch and therefore lowering the luminosity. 

Note that, from the luminosity point of view, the optimal locations for the electron horizontal and vertical focal points may be different, since positron focusing is different in the two transverse directions. 
The optimal electron and positron $\beta^*$ values at their respective focal points may also be different from those listed in Table~\ref{t:vac_par} and used in our simulations. 
Thus, their independent optimization may result in an even higher luminosity than shown in Fig.~\ref{fig:lumi_vs_zec}. 
Nevertheless, optimization of the electron focal point location in Fig.~\ref{fig:lumi_vs_zec} gives a more than two-fold increase in the luminosity to $L = 2.98 \times 10^{34}$~cm$^{-2}$sec$^{-1}$ at $s = 13$~mm over the configuration used in Fig.~\ref{fig:lumi_vs_nslices_0mrad} where the electron and positron optics were set up per Table~\ref{t:vac_par} ignoring the positron focusing effect.

\begin{figure}[t]
   \centering
   \includegraphics*[width=\columnwidth]{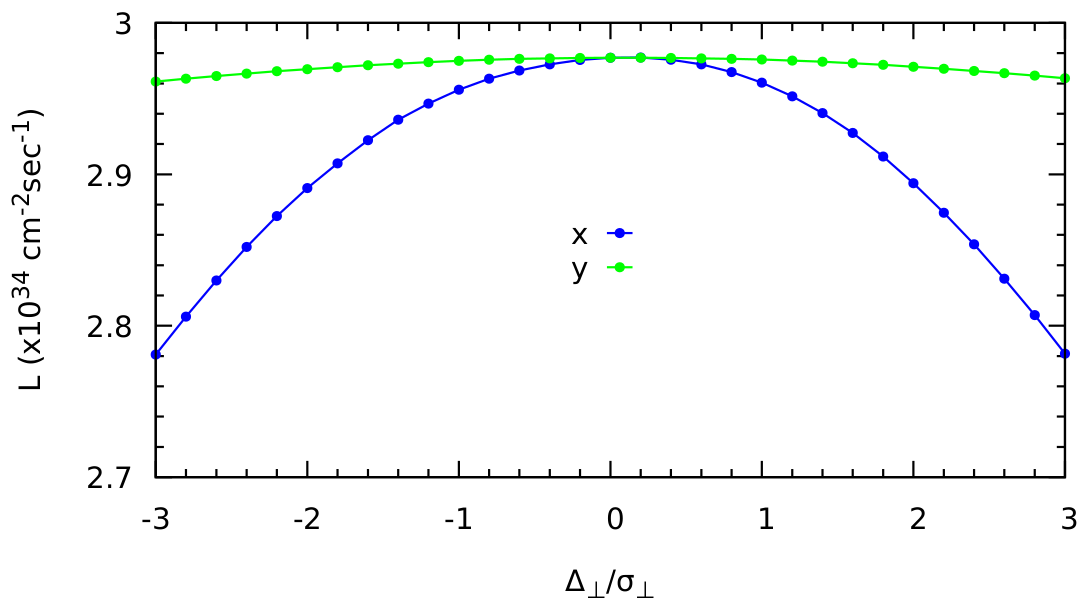}
   \caption{Luminosity as a function of the horizontal (blue) and vertical (green) offsets of the electron and positron beams in units of the nominal rms beam sizes $\sigma_{x,y}^*$ at the IP in the WS case. The rms beam sizes along with other electron and positron beam parameters are listed in Table~\ref{t:vac_par}. The number of positron slices is $m_+ = 6145$. The longitudinal offset of the electron and positron focal points is $\Delta s = 13$~mm.}
   \label{fig:lumi_vs_xec}
\end{figure}

\begin{figure}[t]
   \centering
   \includegraphics*[width=\columnwidth]{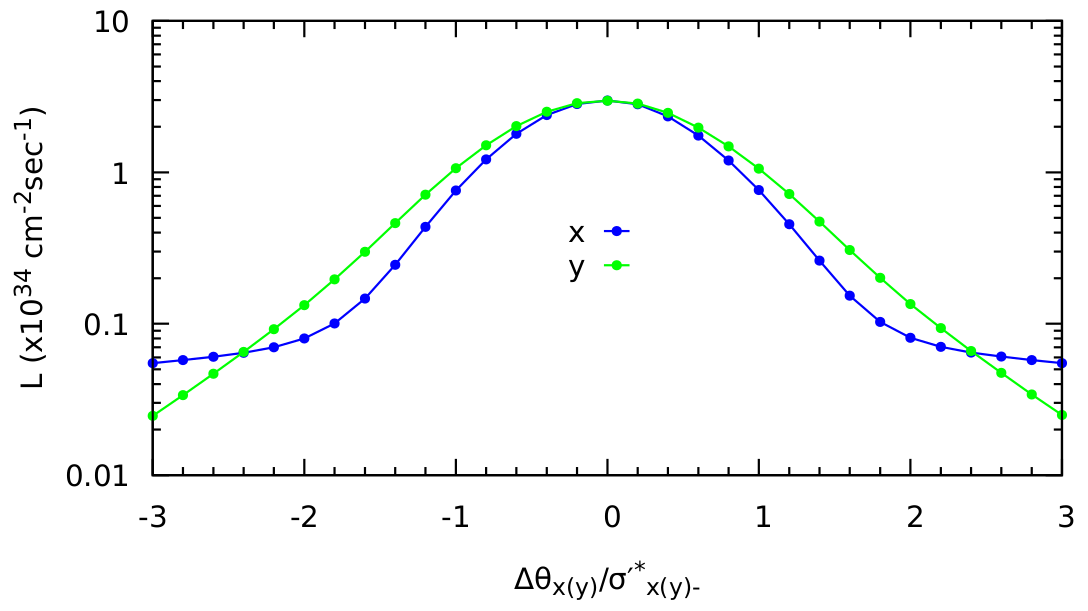}
   \caption{Luminosity as a function of the horizontal (blue) and vertical (green) crossing angles $\Delta \theta_x$ and $\Delta \theta_y$, respectively, of the electron and positron beams normalized to the corresponding nominal rms horizontal ($x$) and vertical ($y$) electron beam divergencies $\sigma_{x,y-}^{\prime *}$ at the IP in the WS case. The rms electron beam angular spreads $\sigma_{x,y-}^{\prime *}$ along with other electron and positron beam parameters are listed in Table~\ref{t:vac_par}. The number of positron slices is $m_+ = 6145$. The longitudinal offset of the electron and positron focal points is $\Delta s = 13$~mm.}
   \label{fig:lumi_vs_xpec}
\end{figure}

Figure~\ref{fig:lumi_vs_xec} demonstrates the sensitivity of the luminosity to the transverse alignment of the electron and positron beams in the WS case. 
As one can see, there is no significant degradation of the luminosity even for $\pm\sigma_\perp$ offsets. 
Note that the horizontal and vertical rms beam sizes in Table~\ref{t:vac_par} are different by a factor of 10. 
This explains the apparently higher sensitivity of the luminosity to the horizontal offset than to the vertical one, since the same number of $\sigma$ in $x$ corresponds to a larger absolute displacement.

We next evaluate the sensitivity of the luminosity to the beam crossing angle. 
The simulation assumed that the nominal beam crossing point is located at $s = 0$, the focal point of the positron beam, while the focal point of the electron beam was still shifted by $\Delta s = 13$~mm. 
The result of this study in the WS case is plotted in Fig.~\ref{fig:lumi_vs_xpec}. 
It shows that the luminosity is flat over an angular range of about $\pm 0.5\sigma_-^{\prime *}$. 
Given the $\sigma_{x,y-}^{\prime *}$ values in Table~\ref{t:vac_par}, this translates into horizontal and vertical angular ranges of about $\pm 10$ and $\pm 5$~mrad, respectively. 
Thus, both the transverse and angular beam alignment requirements do not present a challenge from the luminosity point of view. This is because the electron bunch gets pulled in by the field of the positron bunch.

The algorithm for simulating the beam-beam interaction described above is only valid for head-on collisions. Thus, to properly account for the crossing angle in the simulation of Fig.~\ref{fig:lumi_vs_xpec}, we used the procedure described in Ref.~\cite{Hirata1995}. The basic idea of that procedure is to reduce a collision at an angle to a head-on collision. This is accomplished by Lorentz-transforming both the electron and positron bunches into a frame, which moves with a velocity $\beta = \sin(\Delta_\angle/2)$ in the direction perpendicular to the line bisecting the beam crossing angle in half. The coordinates of each electron and the parameters of each positron slice are transformed accordingly. We do not provide the relevant expressions here for the sake of brevity. They can be found in Ref.~\cite{Hirata1995}. In the new frame, the collision occurs head-on and the above algorithm is applied directly.

\begin{figure*}[t]
\begin{tabular}{ll}
   \centering
   \includegraphics*[width=1\columnwidth]{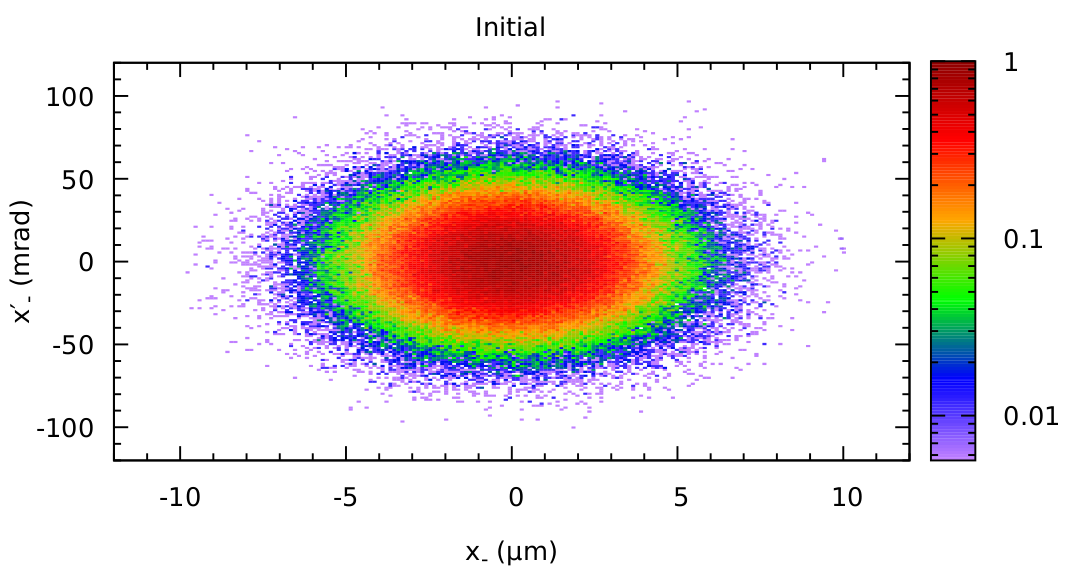}
&
   \includegraphics*[width=1\columnwidth]{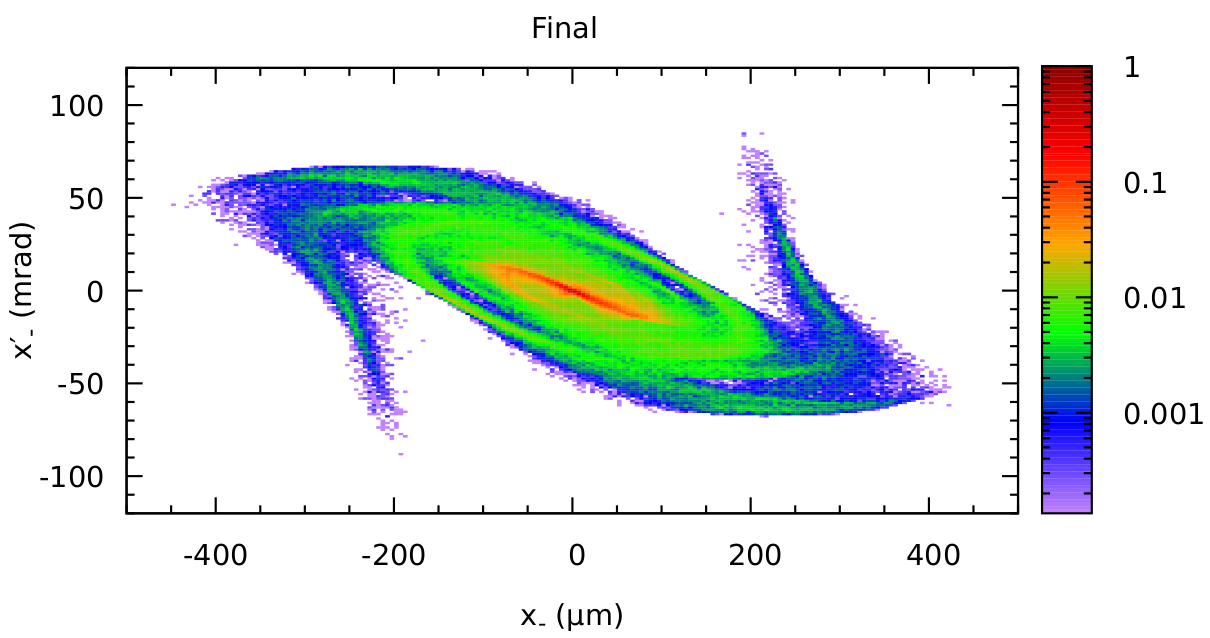} \\
   \centering
   \includegraphics*[width=1\columnwidth]{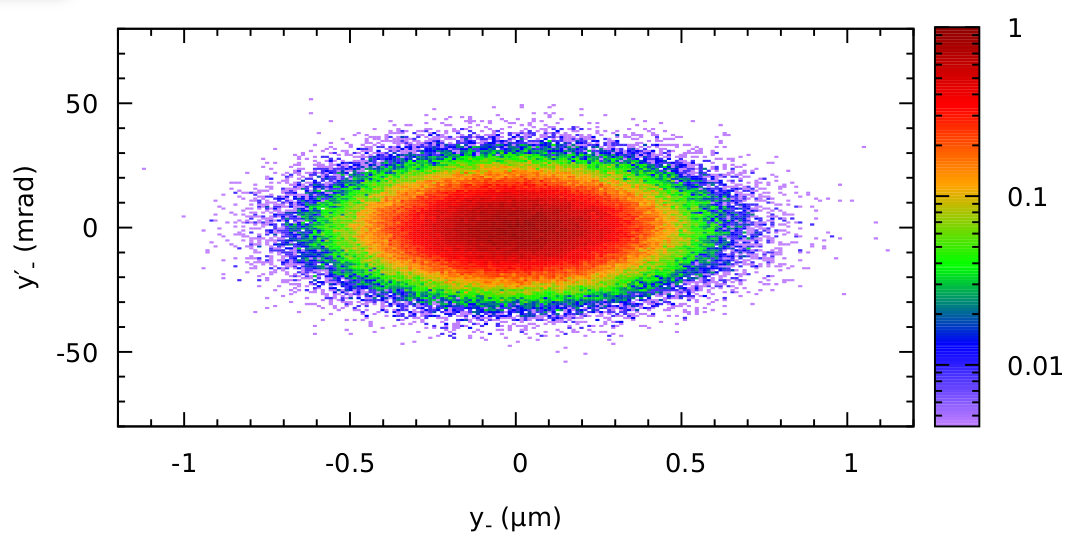}
&
   \includegraphics*[width=1\columnwidth]{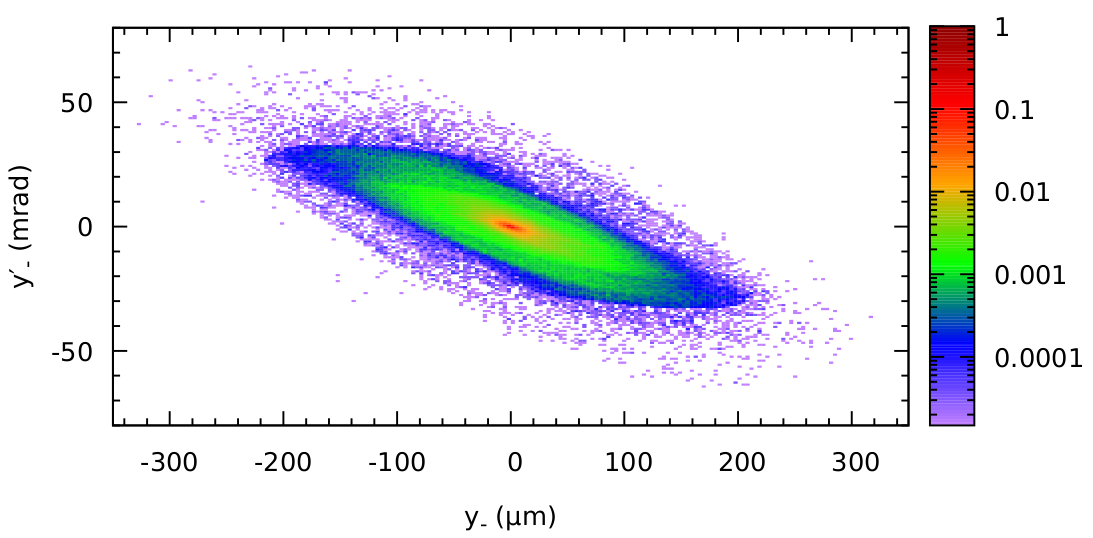} \\
   \centering
   \includegraphics*[width=1\columnwidth]{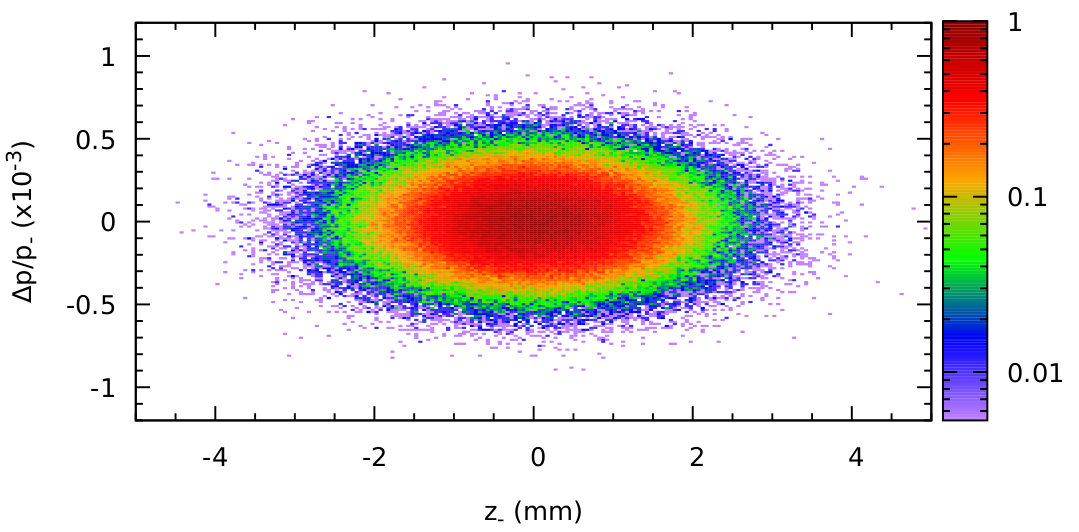}
&
   \includegraphics*[width=1\columnwidth]{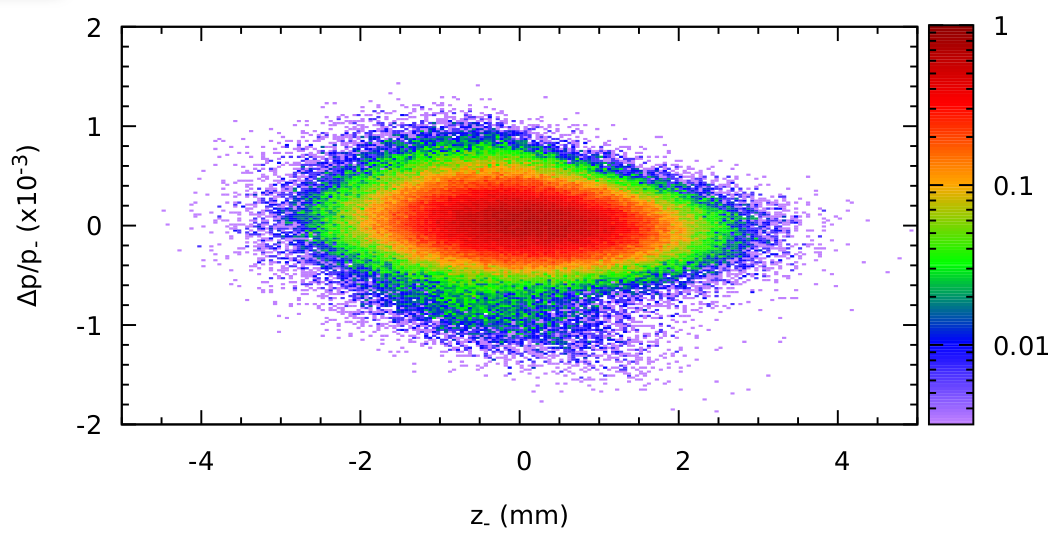}
\end{tabular}
   \caption{Comparison of the horizontal (top), vertical (middle), and longitudinal (bottom) phase-space density distributions of the electron bunch before (left) and after (right) its interaction with the positron bunch in the SS case. In each dimension, the density is normalized to that at the center of the corresponding initial distribution. The electron and positron bunch parameters are listed in Table~\ref{t:vac_par}. Each of the colliding bunches is modeled in BeamBeam3D using about 33.6 million macro-particles on an $x \times y \times z$ grid of $32 \times 32 \times 24576$. The numbers of the electron and positron slices are $m_-=385$ and $m_+ = 3073$, respectively. }
   \label{fig:ele_phsp}
\end{figure*}

Despite the low electron energy, the high fields of the positron bunch and the resulting rapid electron oscillation raise the question of the emitted electron synchrotron radiation power. We numerically evaluate the synchrotron radiation power at the highest luminosity point as
\begin{equation}
P_{\gamma-} = \frac{N_-}{n_-} \frac{C_\gamma E_-^4 }{2\pi} f_b \sum_{i=1}^{n_-} \sum_{j=2}^{m_+} 
\frac{\left| \Delta \vec{p}_{ij-} \right|^2 }{|s_{ij}-s_{ij-1}| p_{zi-}^2} ,
\end{equation}
where $C_\gamma = 8.846 \cdot 10^{-5}$~m/GeV$^3$ is a constant. Our simulation gave a negligible power of $P_{\gamma-} \simeq 11$~mW.

\section{Positron Beam Stability}
One of the greatest concerns for the stability of the stored beam in a linac-ring collider comes from a head-tail type of instability called the kink instability. If the electron and positron bunches are not exactly centered at the interaction point, the electron bunch coherently oscillates through the positron bunch and leaves a longitudinally-dependent imprint on the positron bunch. Under certain conditions, this imprint may get amplified and result in a loss of the positron beam. 

The onset of the kink instability is determined by a threshold $\Lambda_{th}$ on the quantity $\Lambda \equiv D_- \xi_+/\nu_s$ where $\nu_s$ is the positron synchrotron tune. 
At small values of the disruption parameter, when electrons make much less than a full oscillation inside the positron bunch, $\Lambda_{th}$ is a constant of about one~\cite{Hao2013}. 
However, the VAC operates in a unique regime where electrons undergo many oscillations inside the positron bunch. 
There is evidence~\cite{Heifets1990, Li2001, Hao2013} that this situation is more stable and, at large disruption parameters, the threshold scales almost linearly with $D_-$. 

\begin{figure*}[t]
\begin{tabular}{ll}
   \centering
   \includegraphics*[width=1\columnwidth]{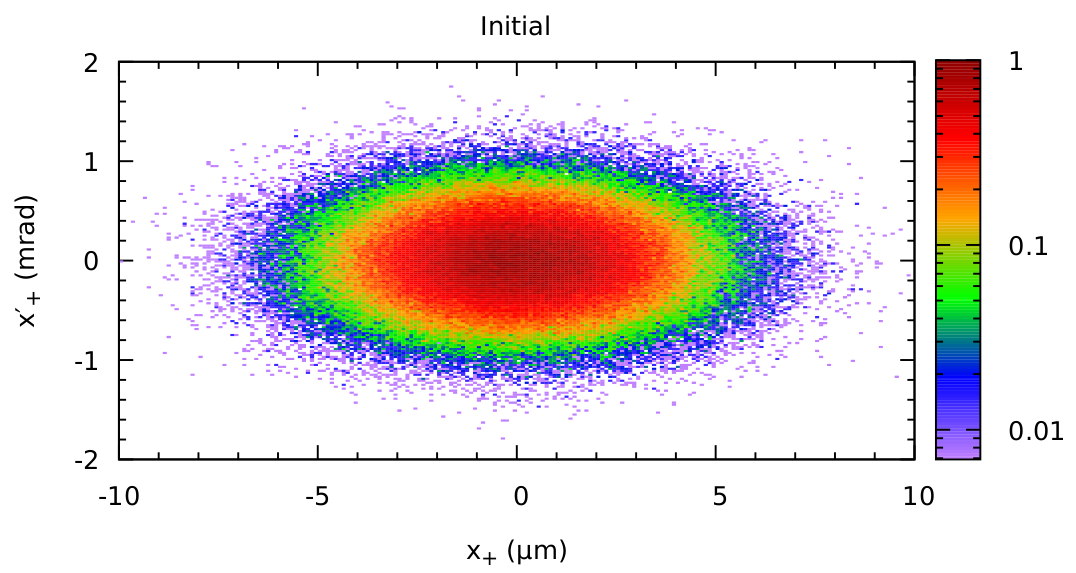}
&
   \includegraphics*[width=1\columnwidth]{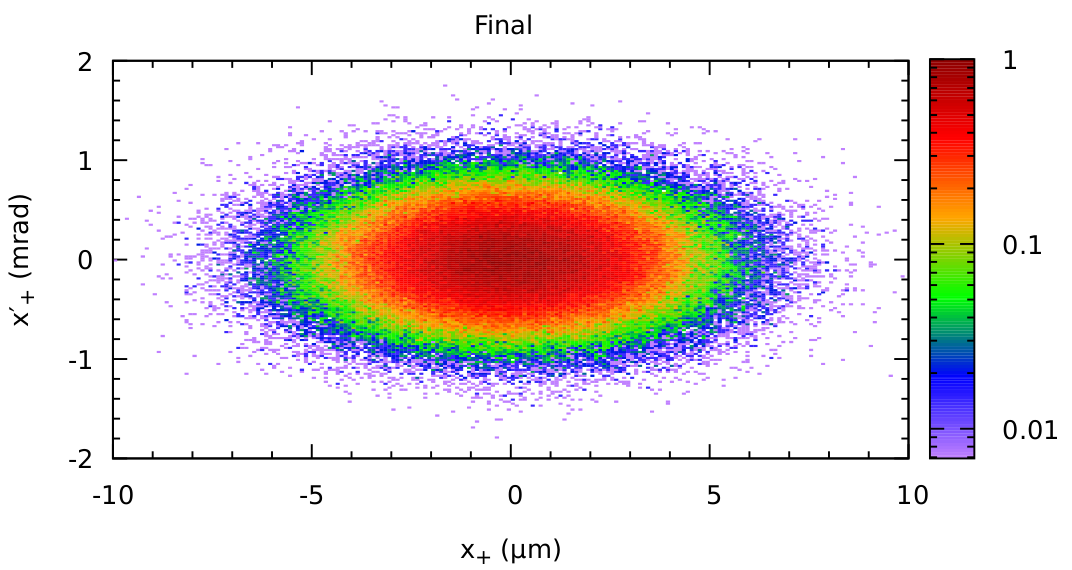} \\
   \centering
   \includegraphics*[width=1\columnwidth]{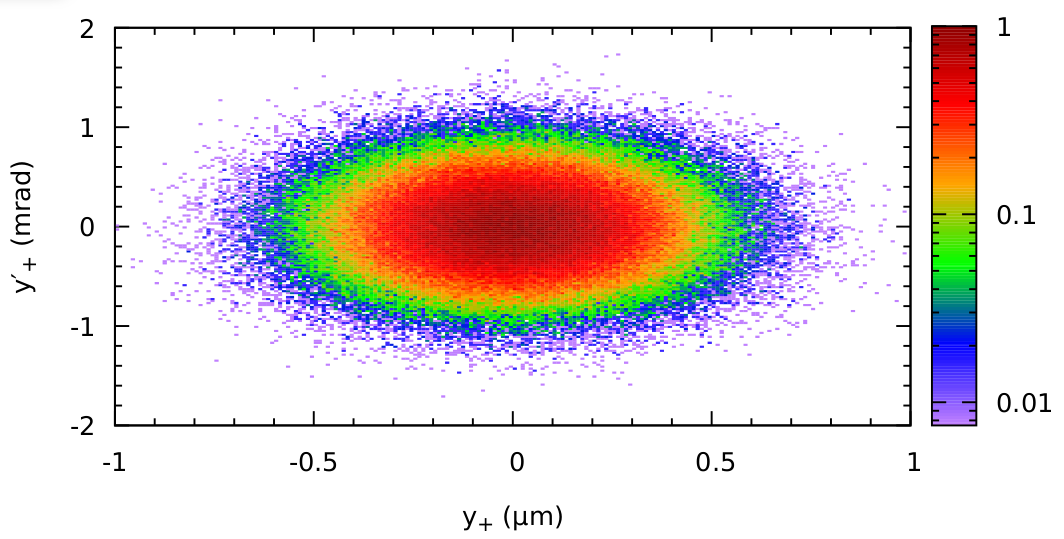}
&
   \includegraphics*[width=1\columnwidth]{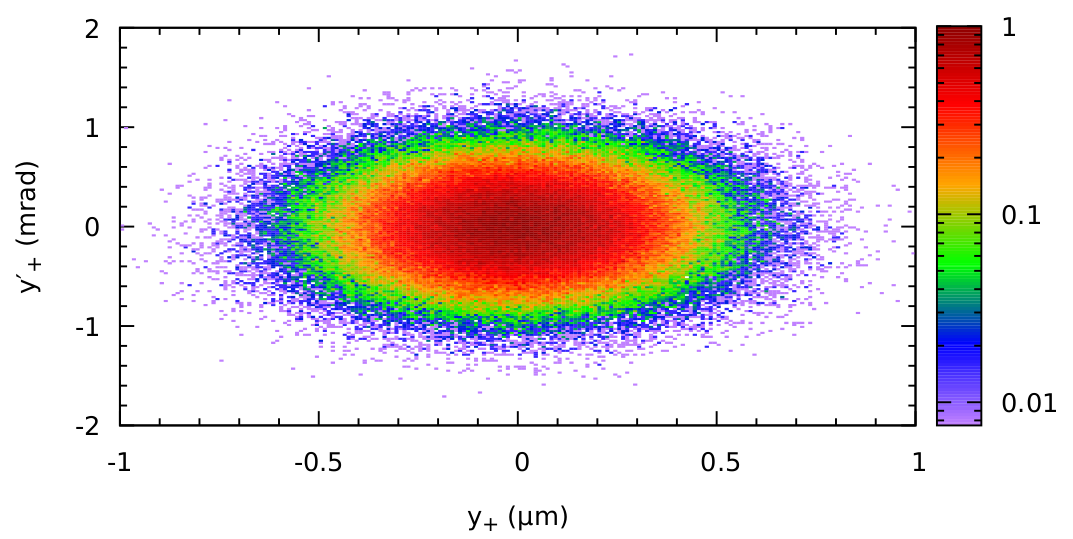} \\
   \centering
   \includegraphics*[width=1\columnwidth]{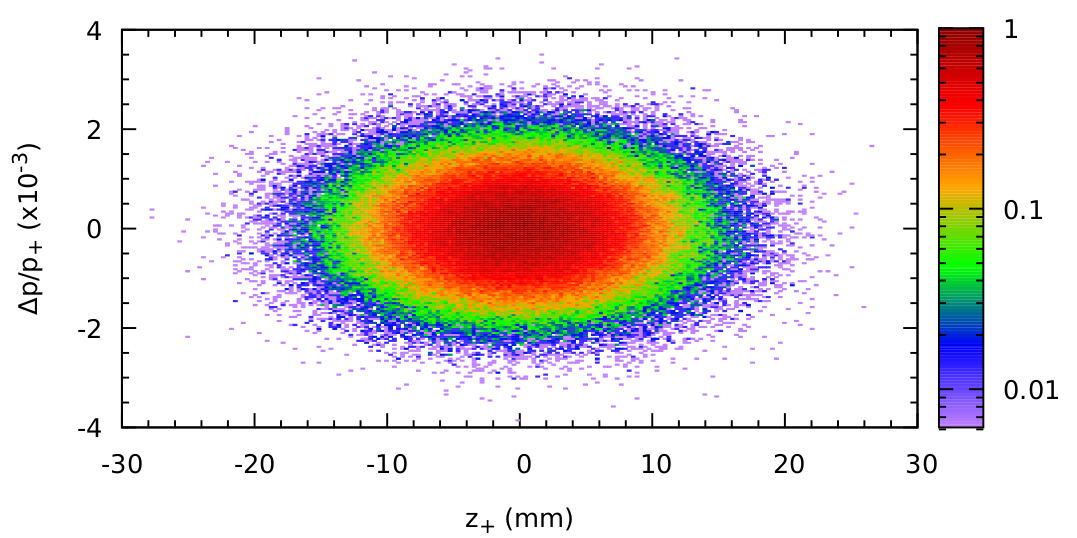}
&
   \includegraphics*[width=1\columnwidth]{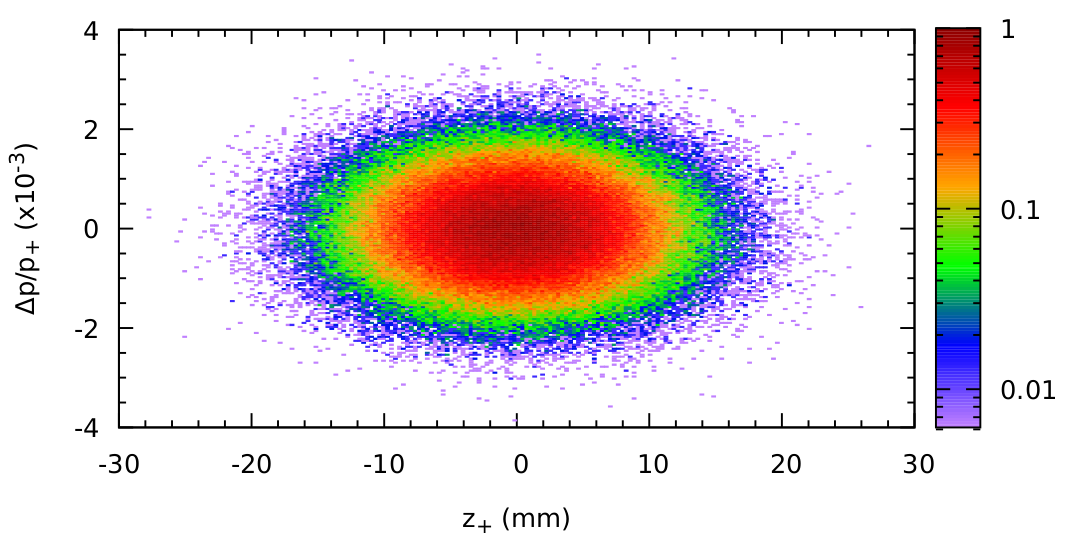}
\end{tabular}
   \caption{Comparison of the horizontal (top), vertical (middle), and longitudinal (bottom) phase-space density distributions of the positron bunch before (left) and after (right) its interaction with the electron bunch in the SS case. In each dimension, the density is normalized to that at the center of the corresponding initial distribution. The electron and positron bunch parameters are listed in Table~\ref{t:vac_par}. Each of the colliding bunches is modeled in BeamBeam3D using about 33.6 million macro-particles on an $x \times y \times z$ grid of $32 \times 32 \times 24576$. The numbers of the electron and positron slices are $m_-=385$ and $m_+ = 3073$, respectively. }
   \label{fig:pos_phsp}
\end{figure*}

A physical interpretation of this behavior is that the electrons leave such a fine ripple on the positron bunch that it gets completely smeared by the synchrotron oscillations in a single turn. 
Thus, the threshold condition becomes a requirement that the ratio $\xi_+/\nu_s$ is less than one. 
The reasoning presented above and the exact threshold value need to be verified by detailed strong-strong simulations.

\begin{figure*}[t]
\begin{tabular}{ll}
   \centering
   \includegraphics*[width=1\columnwidth]{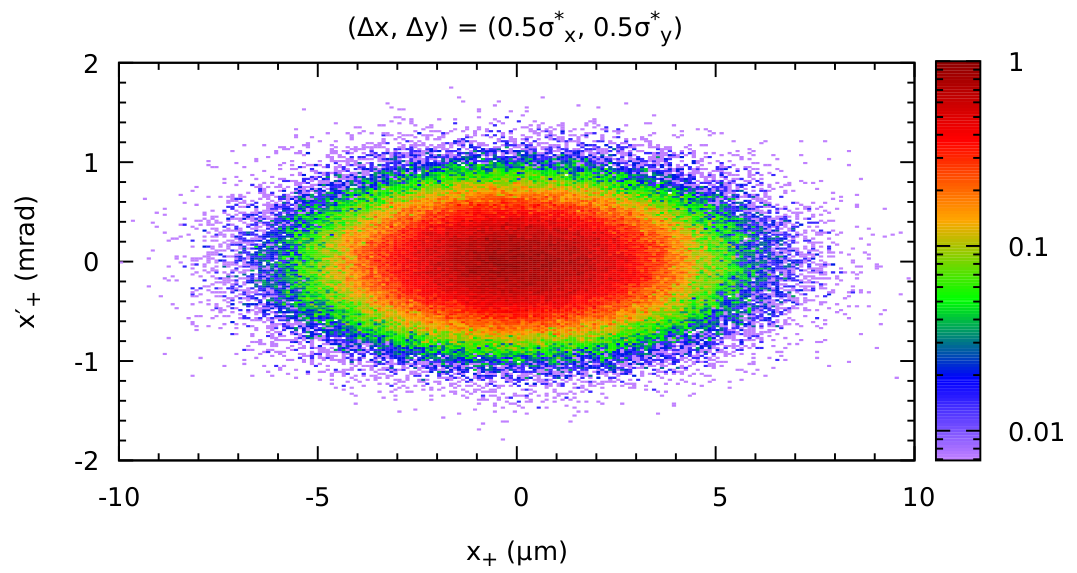}
&
   \includegraphics*[width=1\columnwidth]{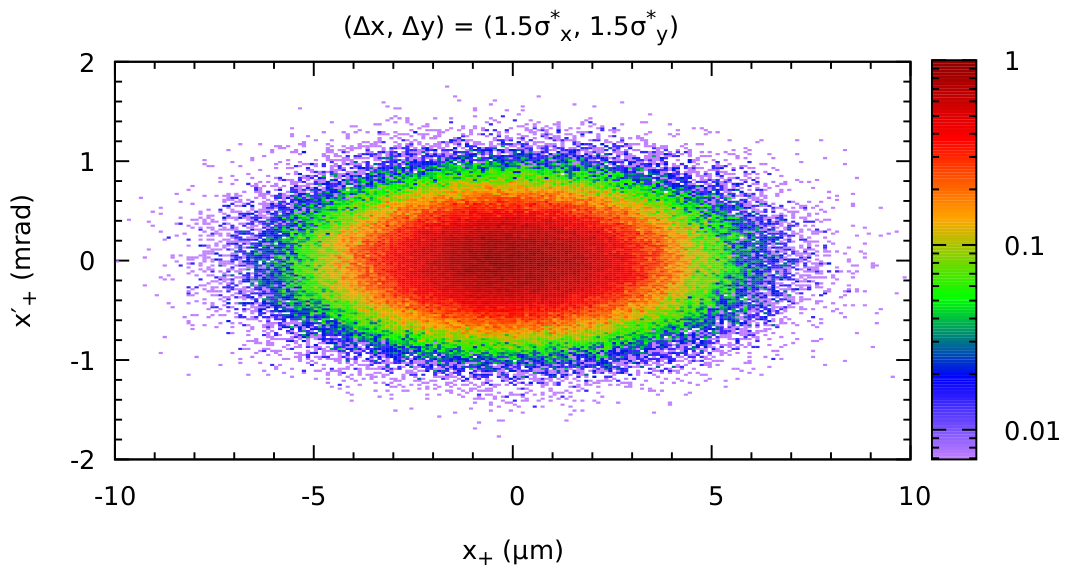} \\
   \centering
   \includegraphics*[width=1\columnwidth]{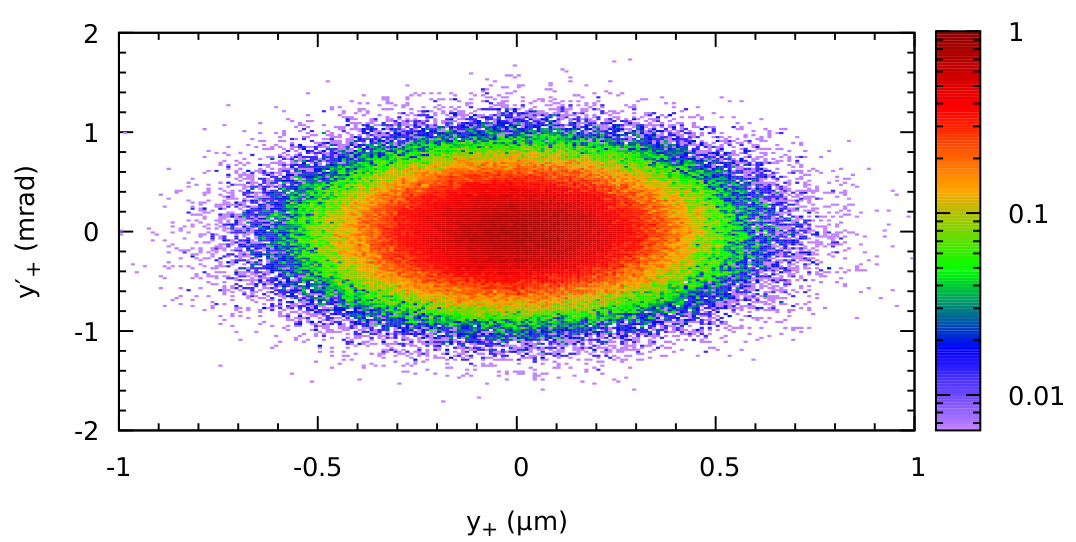}
&
   \includegraphics*[width=1\columnwidth]{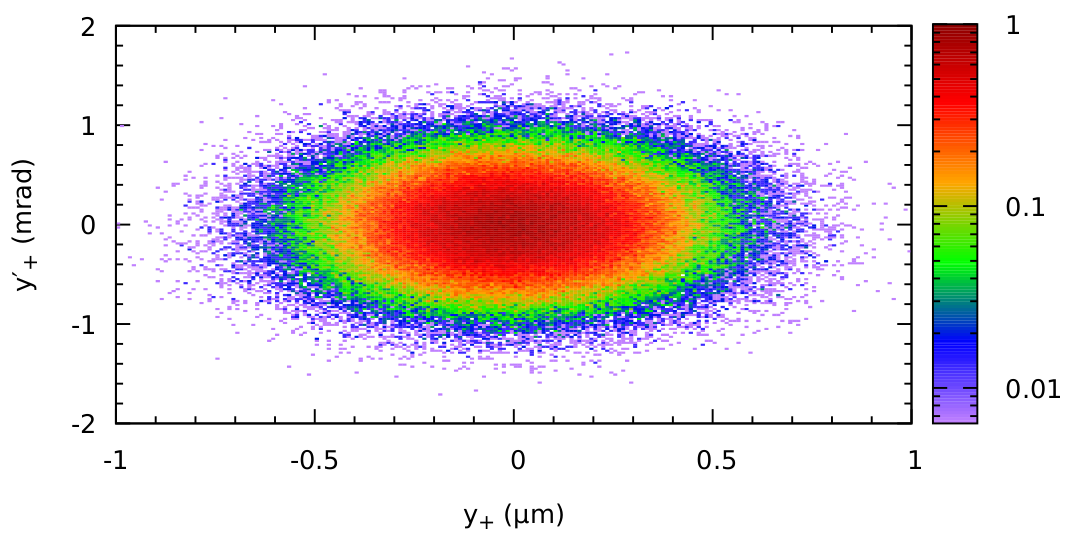} \\
   \centering
   \includegraphics*[width=1\columnwidth]{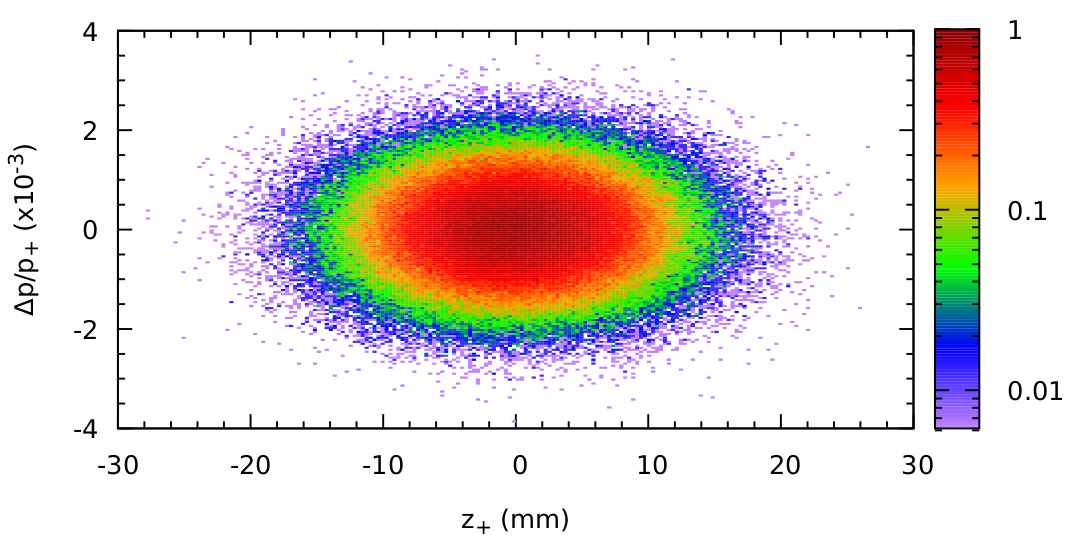}
&
   \includegraphics*[width=1\columnwidth]{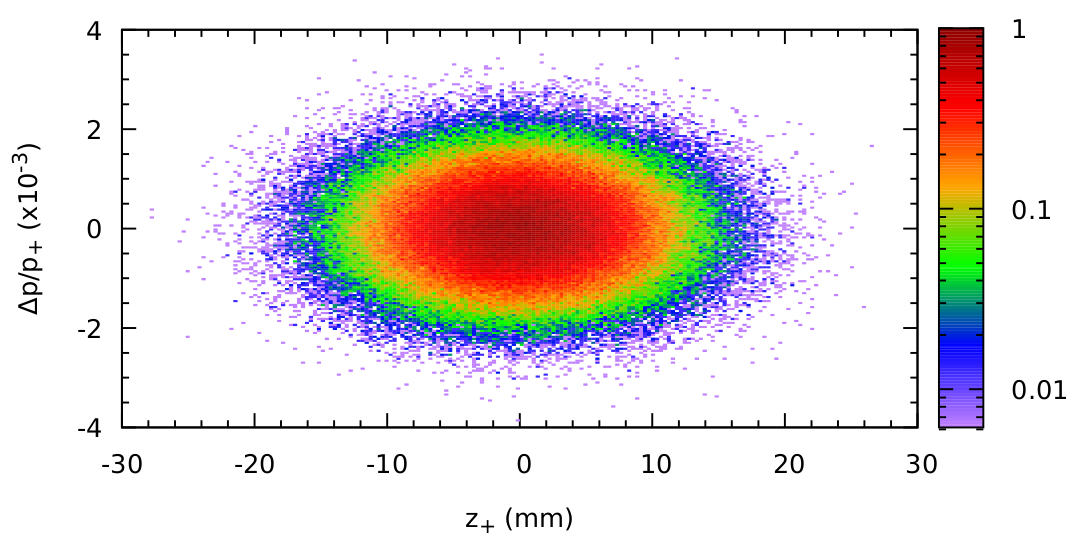}
\end{tabular}
   \caption{Final horizontal (top), vertical (middle), and longitudinal (bottom) phase-space density distributions of the positron bunch after its collision with an electron bunch for the cases when the incident electron bunch is initially offset by $(\Delta x, \Delta y)$ of $(0.5 \sigma_x^*, 0.5 \sigma_y^*)$ (left) and $(1.5 \sigma_x^*, 1.5 \sigma_y^*)$ (right). The density in each plot of each dimension is normalized to that at the center of the corresponding initial distribution shown on the left-hand side of Fig.~\ref{fig:pos_phsp} . The collision is simulated using BeamBeam3D in the SS mode. The electron and positron bunch parameters are listed in Table~\ref{t:vac_par}. Each of the colliding bunches is modeled using about 33.6 million macro-particles on an $x \times y \times z$ grid of $32 \times 32 \times 24576$. The numbers of the electron and positron slices are $m_-=385$ and $m_+ = 3073$, respectively. }
   \label{fig:pos_diff_ele_off}
\end{figure*}

To further justify our above reasoning, we compare the electron and positron bunch phase-space distributions before and after a single interaction in Figs.~\ref{fig:ele_phsp} and \ref{fig:pos_phsp}, respectively. The bunch interaction is modeled in BeamBeam3D using the SS mode. The electron and positron bunch parameters are listed in Table~\ref{t:vac_par}. Each of the colliding bunches is represented by about 33.6 million macro-particles on an $x \times y \times z$ grid of $32 \times 32 \times 24576$. The numbers of the electron and positron slices are $m_-=385$ and $m_+ = 3073$, respectively. The electron focal point is not shifted. The results in Figs.~\ref{fig:ele_phsp} and \ref{fig:pos_phsp} correspond to the $m_-=385$, $m_+ = 3073$ SS luminosity point of Fig.~\ref{fig:lumi_vs_nslices_0mrad}. Note that both the initial and final distributions of both bunches are shown projected to the IP. While Figs.~\ref{fig:ele_phsp} exhibits drastic distortion of the electron bunch by the single beam-beam interaction, there is no discernible perturbation of the shape of the positron bunch.

To illustrate the positron bunch stability against position jitter of the electron bunch, we study the effect of the transverse offset of an electron bunch on the final positron phase-space distribution after a single collision. The simulation setup parameters are the same as described above for Figs.~\ref{fig:ele_phsp} and \ref{fig:pos_phsp}. The electron bunch is offset simultaneously in both $x$ and $y$ by $(\Delta x, \Delta y)$ of $(0.5 \sigma_x^*, 0.5 \sigma_y^*)$ and $(1.5 \sigma_x^*, 1.5 \sigma_y^*)$. The final positron distributions corresponding to these offsets are shown on the left- and right-hand sides of Fig.~\ref{fig:pos_diff_ele_off}, respectively. Their comparison to the perfectly aligned case on the right-hand side of Fig.~\ref{fig:pos_phsp} shows no visually distinguishable difference between any of these cases. Note that, while the different data sets appear the same, this is not a plotting error; the data points are different by up to several parts in $10^4$. While reassuring, this is not a conclusive demonstration of a long-term positron bunch stability. It still needs to be verified by more detailed multi-turn simulations, which go beyond the scope of this paper.

Note that, in the above simulations, we assumed the electron bunch to have a Gaussian distribution. It may not be the case for a linac beam. However, from a conceptual design point of view, the exact shape of the electron distribution should not make a dramatic difference. It may modify the beam-beam kick experienced by the positron bunch but, as shown above, it is so small to begin with that it is unlikely that its exact profile can make a drastic difference. The effects of the bunch shape mismatch and of the particular electron bunch distribution need to be studied in conjunction with a specific linac design when the proposed project reaches that level of detail in its development. 

\section{Interaction region}
A dark matter particle in the $e^+e^- \rightarrow \gamma + A'$ process could be discovered using the missing mass method with the observed photon. 
In the center of mass frame, photons are strongly picked along the beam direction, see Ref.~\cite{Fayet2007}. 
Due to the energy asymmetry in the colliding beams, the collision products are strongly boosted forward along the outgoing positron beam direction, so that photons make a small angle with it. 
Thus, photon detection primarily requires an experimental detector that has an acceptance in solid angle around the forward going positron beam. 
This makes a VAC detector region design fairly straightforward.

\begin{figure}[t]
   \centering
   \includegraphics*[width=\columnwidth]{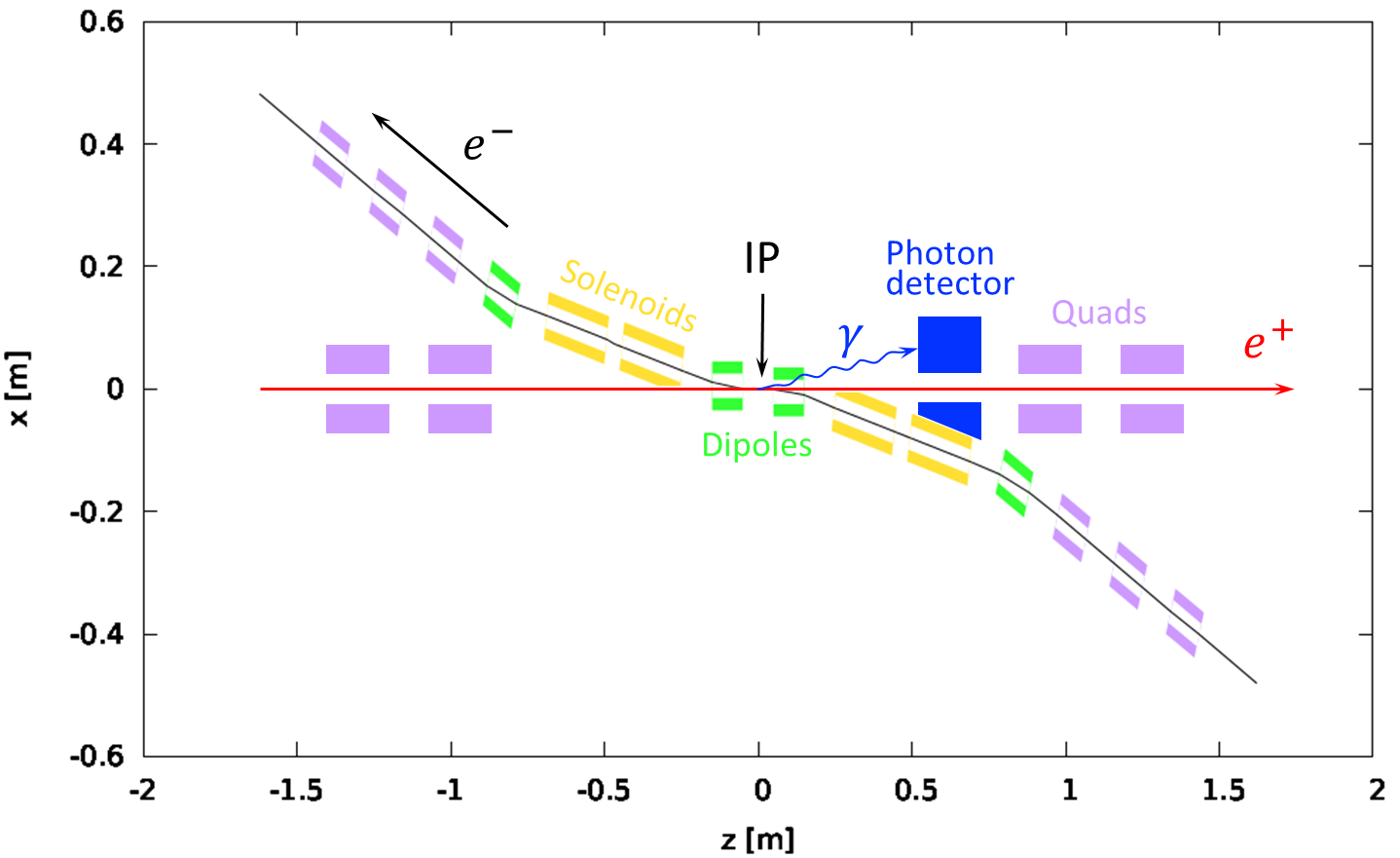}
   \caption{Possible schematic of a VAC interaction region showing its main components.}
   \label{fig:ir_layout}
\end{figure}

\begin{figure}[t]
   \centering
   \includegraphics*[width=\columnwidth]{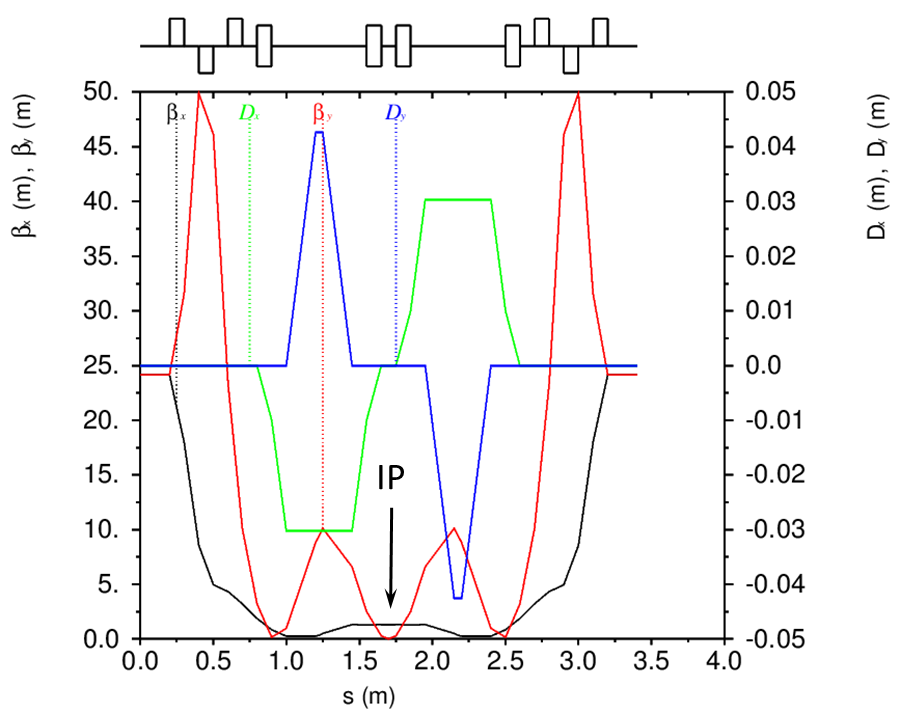}
   \caption{Magnetic optics concept of a VAC interaction region for the layout shown in Fig.~\ref{fig:ir_layout}.}
   \label{fig:ir_optics}
\end{figure}

A possible schematic of a VAC interaction region is shown in Fig.~\ref{fig:ir_layout}. 
The energy difference between the beams makes it easy to bring them together before and separate them after a collision. 
A 10~cm long 0.067~T dipole bends the electron beam by 200~mrad while the positron is deflected by only 0.5~mrad. As shown in Fig.~\ref{fig:ir_layout}, there are two such dipoles on each side of the IP. 
The pairs of dipoles before and after the IP are arranged as Double-Bend Achromats (DBAs) to prevent creation of dispersion at the IP as shown in Fig.~\ref{fig:ir_optics}. 
The overall electron IR geometry has a dogleg configuration. 

An electron beam in this energy range can be effectively focused by solenoids. Thus, the necessary betatron phase advance between the two dipoles in each DBA is produced by a pair of solenoids. The solenoids in each pair are powered with opposite polarities to avoid introducing transverse coupling while still providing full beam focusing. The solenoids also serve as a part of the final focusing system. 

The IR optics shown in Fig.~\ref{fig:ir_optics} assumes that the electron beam has already been converted from a round magnetized beam to an uncoupled flat beam with equal horizontal and vertical $\beta$ functions. Such a beam can be readily transported using axially-symmetric focusing. However, the VAC parameters call for highly asymmetric $\beta$ functions at the IP. Thus, quadrupole triplets upstream and downstream of the respective DBAs complement the solenoid focusing to provide the necessary beam parameters at the IP. 

The positron beam parameters at the IP are already consistent with those at SuperKEK-B. 
Therefore, we do not discuss the positron IR design in detail. 
There are many beam dynamics aspects associated with an IR design for a collider ring but the performance necessary for the VAC has already been demonstrated at SuperKEK-B. 
Assuming that the projected reduction in the horizontal positron beam emittance can be reached, it should be straightforward to reduce the horizontal $\beta^*$ from the SuperKEK-B value, as is desirable for a VAC. 
The VAC physics requirements allow for a much simpler machine-detector interface compared to that of Belle II. As shown in Fig.~\ref{fig:ir_layout}, the central part of the VAC detector could be on the order of 0.5~m in length while the entire IR has a scale of a few meters.

Given the small beam size and the associated large angular beam divergence at the IP, the presence of chromatic and spherical beam aberrations may cause smear of the beam spot size at the focal point leading to reduction in the luminosity. 
Proper measures must be taken to prevent this effect. A detailed discussion of this topic is beyond the scope of this paper. 
However, compensation for the non-linear beam smear at the IP is a well-understood subject and appropriate techniques have been developed~\cite{Raimondi2001, Morozov2013}. 

\section{Summary}
We describe our concept of a Very-Asymmetric lepton Collider for a Dark Matter search. Even though further studies are needed, we show that it is feasible to reach a high luminosity of the order of $10^{34}$~cm$^{-2}$s$^{-1}$ in such a collider in the center-of-mass energy range well below~1 GeV as necessitated by the relevant physics. 
Such a high luminosity performance relies on the high quality and high current of a stored positron beam and those of a magnetized electron beam from a linac. Only reasonable extrapolations are assumed from the current state of the art. We demonstrate that, with the mutual beam-beam interaction accounted for, the luminosity is a few times $10^{34}$~cm$^{-2}$s$^{-1}$. A more systematic and more comprehensive optimization of the beam parameters than presented here may provide an even higher luminosity. We briefly discuss the question of stability of the stored positron beam and conclude that there may not be fundamental issues. The interaction region design is straightforward given the VAC physics goals. Overall, such a collider on a modest energy scale may open access to a new field of research with a potential of discovery of physics beyond the Standard Model.

\begin{acknowledgments}
We appreciate and are thankful to Y.~Derbenev of Jefferson Lab for fruitful discussions. We also thank J. Qiang of Berkeley Lab for his help with setup of BeamBeam3D simulations.  
This manuscript has been authored in part by UT-Battelle, LLC, under Contract No. DE-AC05-00OR22725 and Jefferson Science Associates, LLC under Contract No. DE-AC05-06OR23177 with the US Department of Energy (DOE). 
The publisher acknowledges the US government license to provide public access under the DOE Public Access Plan (http://energy.gov/downloads/doe-public-access-plan).
\end{acknowledgments}

\bibliographystyle{apsrev4-1}
\bibliography{VAC.bib}

\begin{thebibliography}{40}%
\makeatletter
\providecommand \@ifxundefined [1]{%
 \@ifx{#1\undefined}
}%
\providecommand \@ifnum [1]{%
 \ifnum #1\expandafter \@firstoftwo
 \else \expandafter \@secondoftwo
 \fi
}%
\providecommand \@ifx [1]{%
 \ifx #1\expandafter \@firstoftwo
 \else \expandafter \@secondoftwo
 \fi
}%
\providecommand \natexlab [1]{#1}%
\providecommand \enquote  [1]{``#1''}%
\providecommand \bibnamefont  [1]{#1}%
\providecommand \bibfnamefont [1]{#1}%
\providecommand \citenamefont [1]{#1}%
\providecommand \href@noop [0]{\@secondoftwo}%
\providecommand \href [0]{\begingroup \@sanitize@url \@href}%
\providecommand \@href[1]{\@@startlink{#1}\@@href}%
\providecommand \@@href[1]{\endgroup#1\@@endlink}%
\providecommand \@sanitize@url [0]{\catcode `\\12\catcode `\$12\catcode
  `\&12\catcode `\#12\catcode `\^12\catcode `\_12\catcode `\%12\relax}%
\providecommand \@@startlink[1]{}%
\providecommand \@@endlink[0]{}%
\providecommand \url  [0]{\begingroup\@sanitize@url \@url }%
\providecommand \@url [1]{\endgroup\@href {#1}{\urlprefix }}%
\providecommand \urlprefix  [0]{URL }%
\providecommand \Eprint [0]{\href }%
\providecommand \doibase [0]{http://dx.doi.org/}%
\providecommand \selectlanguage [0]{\@gobble}%
\providecommand \bibinfo  [0]{\@secondoftwo}%
\providecommand \bibfield  [0]{\@secondoftwo}%
\providecommand \translation [1]{[#1]}%
\providecommand \BibitemOpen [0]{}%
\providecommand \bibitemStop [0]{}%
\providecommand \bibitemNoStop [0]{.\EOS\space}%
\providecommand \EOS [0]{\spacefactor3000\relax}%
\providecommand \BibitemShut  [1]{\csname bibitem#1\endcsname}%
\let\auto@bib@innerbib\@empty
\bibitem [{\citenamefont {Arkani-Hamed}\ \emph {et~al.}(2009)\citenamefont
  {Arkani-Hamed}, \citenamefont {Finkbeiner}, \citenamefont {Slatyer},\ and\
  \citenamefont {Weiner}}]{Arkani-Hamed}%
  \BibitemOpen
  \bibfield  {author} {\bibinfo {author} {\bibfnamefont {N.}~\bibnamefont
  {Arkani-Hamed}}, \bibinfo {author} {\bibfnamefont {D.~P.}\ \bibnamefont
  {Finkbeiner}}, \bibinfo {author} {\bibfnamefont {T.~R.}\ \bibnamefont
  {Slatyer}}, \ and\ \bibinfo {author} {\bibfnamefont {N.}~\bibnamefont
  {Weiner}},\ }\href {\doibase 10.1103/PhysRevD.79.015014} {\bibfield
  {journal} {\bibinfo  {journal} {Phys. Rev. D}\ }\textbf {\bibinfo {volume}
  {79}},\ \bibinfo {pages} {015014} (\bibinfo {year} {2009})}\BibitemShut
  {NoStop}%
\bibitem [{\citenamefont {Bjorken}\ \emph {et~al.}(2009)\citenamefont
  {Bjorken}, \citenamefont {Essig}, \citenamefont {Schuster},\ and\
  \citenamefont {Toro}}]{BEST}%
  \BibitemOpen
  \bibfield  {author} {\bibinfo {author} {\bibfnamefont {J.~D.}\ \bibnamefont
  {Bjorken}}, \bibinfo {author} {\bibfnamefont {R.}~\bibnamefont {Essig}},
  \bibinfo {author} {\bibfnamefont {P.}~\bibnamefont {Schuster}}, \ and\
  \bibinfo {author} {\bibfnamefont {N.}~\bibnamefont {Toro}},\ }\href {\doibase
  10.1103/PhysRevD.80.075018} {\bibfield  {journal} {\bibinfo  {journal} {Phys.
  Rev. D}\ }\textbf {\bibinfo {volume} {80}},\ \bibinfo {pages} {075018}
  (\bibinfo {year} {2009})}\BibitemShut {NoStop}%
\bibitem [{\citenamefont {Pospelov}(2009)}]{Pospelov}%
  \BibitemOpen
  \bibfield  {author} {\bibinfo {author} {\bibfnamefont {M.}~\bibnamefont
  {Pospelov}},\ }\href {\doibase 10.1103/PhysRevD.80.095002} {\bibfield
  {journal} {\bibinfo  {journal} {Phys. Rev. D}\ }\textbf {\bibinfo {volume}
  {80}},\ \bibinfo {pages} {095002} (\bibinfo {year} {2009})}\BibitemShut
  {NoStop}%
\bibitem [{\citenamefont {Izaguirre}\ \emph {et~al.}(2014)\citenamefont
  {Izaguirre}, \citenamefont {Krnjaic}, \citenamefont {Schuster},\ and\
  \citenamefont {Toro}}]{Toro}%
  \BibitemOpen
  \bibfield  {author} {\bibinfo {author} {\bibfnamefont {E.}~\bibnamefont
  {Izaguirre}}, \bibinfo {author} {\bibfnamefont {G.}~\bibnamefont {Krnjaic}},
  \bibinfo {author} {\bibfnamefont {P.}~\bibnamefont {Schuster}}, \ and\
  \bibinfo {author} {\bibfnamefont {N.}~\bibnamefont {Toro}},\ }\href {\doibase
  10.1103/PhysRevD.90.014052} {\bibfield  {journal} {\bibinfo  {journal} {Phys.
  Rev. D}\ }\textbf {\bibinfo {volume} {90}},\ \bibinfo {pages} {014052}
  (\bibinfo {year} {2014})}\BibitemShut {NoStop}%
\bibitem [{\citenamefont {Abe}\ \emph {et~al.}(2013)\citenamefont {Abe} \emph
  {et~al.}}]{Abe2013}%
  \BibitemOpen
  \bibfield  {author} {\bibinfo {author} {\bibfnamefont {T.}~\bibnamefont
  {Abe}} \emph {et~al.},\ }\href {\doibase 10.1093/ptep/pts102} {\bibfield
  {journal} {\bibinfo  {journal} {Prog. Theor. Exp. Phys.}\ }\textbf {\bibinfo
  {volume} {2013}},\ \bibinfo {pages} {03A001} (\bibinfo {year}
  {2013})}\BibitemShut {NoStop}%
\bibitem [{\citenamefont {Ohnishi}\ \emph {et~al.}(2013)\citenamefont {Ohnishi}
  \emph {et~al.}}]{Ohnishi2013}%
  \BibitemOpen
  \bibfield  {author} {\bibinfo {author} {\bibfnamefont {Y.}~\bibnamefont
  {Ohnishi}} \emph {et~al.},\ }\href {\doibase
  https://doi.org/10.1093/ptep/pts083} {\bibfield  {journal} {\bibinfo
  {journal} {Prog. Theor. Exp. Phys.}\ }\textbf {\bibinfo {volume} {2013}},\
  \bibinfo {pages} {03A001} (\bibinfo {year} {2013})}\BibitemShut {NoStop}%
\bibitem [{\citenamefont {{{KEK} web page}}(2022)}]{KEK}%
  \BibitemOpen
  \bibfield  {author} {\bibinfo {author} {\bibnamefont {{{KEK} web page}}},\
  }\href {https://www.kek.jp/en/newsroom/2020/06/26/1400/} {\enquote {\bibinfo
  {title} {Status of the super {KEK} {B} factory},}\ } (\bibinfo {year}
  {2022})\BibitemShut {NoStop}%
\bibitem [{\citenamefont {Seeman}(2001)}]{Seeman2001}%
  \BibitemOpen
  \bibfield  {author} {\bibinfo {author} {\bibfnamefont {J.~T.}\ \bibnamefont
  {Seeman}},\ }in\ \href {\doibase
  https://aip.scitation.org/doi/abs/10.1063/1.1420415} {\emph {\bibinfo
  {booktitle} {AIP Conf. Proc.}}},\ Vol.\ \bibinfo {volume} {592}\ (\bibinfo
  {year} {2001})\ p.\ \bibinfo {pages} {163}\BibitemShut {NoStop}%
\bibitem [{\citenamefont {Wojtsekhowski}\ \emph {et~al.}(2017)\citenamefont
  {Wojtsekhowski}, \citenamefont {Morozov},\ and\ \citenamefont
  {Derbenev}}]{BW_2017}%
  \BibitemOpen
  \bibfield  {author} {\bibinfo {author} {\bibfnamefont {B.}~\bibnamefont
  {Wojtsekhowski}}, \bibinfo {author} {\bibfnamefont {V.~S.}\ \bibnamefont
  {Morozov}}, \ and\ \bibinfo {author} {\bibfnamefont {Y.~S.}\ \bibnamefont
  {Derbenev}},\ }\href@noop {} {\enquote {\bibinfo {title} {Very {A}symmetric
  {C}ollider for {D}ark {M}atter {S}earch below 1 {GeV}},}\ } (\bibinfo {year}
  {2017}),\ \Eprint {http://arxiv.org/abs/1705.00051} {arXiv:1705.00051}
  \BibitemShut {NoStop}%
\bibitem [{\citenamefont {Wojtsekhowski}(2009)}]{BW_2009}%
  \BibitemOpen
  \bibfield  {author} {\bibinfo {author} {\bibfnamefont {B.}~\bibnamefont
  {Wojtsekhowski}},\ }\href
  {https://www.jlab.org/conferences/JPOS09/program.html} {\enquote {\bibinfo
  {title} {U-boson search with positrons},}\ } (\bibinfo {year} {2009}),\
  \Eprint {http://arxiv.org/abs/0906.5265} {arXiv:0906.5265} \BibitemShut
  {NoStop}%
\bibitem [{\citenamefont {Wojtsekhowski}\ \emph {et~al.}(2018)\citenamefont
  {Wojtsekhowski} \emph {et~al.}}]{BW_2018}%
  \BibitemOpen
  \bibfield  {author} {\bibinfo {author} {\bibfnamefont {B.}~\bibnamefont
  {Wojtsekhowski}} \emph {et~al.},\ }\href {\doibase
  https://doi.org/10.1088/1748-0221/13/02/p02021} {\bibfield  {journal}
  {\bibinfo  {journal} {Journal of Instrumentation}\ }\textbf {\bibinfo
  {volume} {13}},\ \bibinfo {pages} {P02021} (\bibinfo {year}
  {2018})}\BibitemShut {NoStop}%
\bibitem [{\citenamefont {Burov}\ \emph {et~al.}(2000)\citenamefont {Burov},
  \citenamefont {Nagaitsev}, \citenamefont {Shemyakin},\ and\ \citenamefont
  {Derbenev}}]{Burov2000}%
  \BibitemOpen
  \bibfield  {author} {\bibinfo {author} {\bibfnamefont {A.}~\bibnamefont
  {Burov}}, \bibinfo {author} {\bibfnamefont {S.}~\bibnamefont {Nagaitsev}},
  \bibinfo {author} {\bibfnamefont {A.}~\bibnamefont {Shemyakin}}, \ and\
  \bibinfo {author} {\bibfnamefont {Y.}~\bibnamefont {Derbenev}},\ }\href
  {\doibase 10.1103/PhysRevSTAB.3.094002} {\bibfield  {journal} {\bibinfo
  {journal} {Phys. Rev. ST Accel. Beams}\ }\textbf {\bibinfo {volume} {3}},\
  \bibinfo {pages} {094002} (\bibinfo {year} {2000})}\BibitemShut {NoStop}%
\bibitem [{\citenamefont {Brinkmann}\ \emph {et~al.}(2001)\citenamefont
  {Brinkmann}, \citenamefont {Derbenev},\ and\ \citenamefont
  {Fl\"ottmann}}]{Brinkmann2001}%
  \BibitemOpen
  \bibfield  {author} {\bibinfo {author} {\bibfnamefont {R.}~\bibnamefont
  {Brinkmann}}, \bibinfo {author} {\bibfnamefont {Y.}~\bibnamefont {Derbenev}},
  \ and\ \bibinfo {author} {\bibfnamefont {K.}~\bibnamefont {Fl\"ottmann}},\
  }\href {\doibase 10.1103/PhysRevSTAB.4.053501} {\bibfield  {journal}
  {\bibinfo  {journal} {Phys. Rev. ST Accel. Beams}\ }\textbf {\bibinfo
  {volume} {4}},\ \bibinfo {pages} {053501} (\bibinfo {year}
  {2001})}\BibitemShut {NoStop}%
\bibitem [{\citenamefont {Burov}\ \emph {et~al.}(2002)\citenamefont {Burov},
  \citenamefont {Nagaitsev},\ and\ \citenamefont {Derbenev}}]{Burov2002}%
  \BibitemOpen
  \bibfield  {author} {\bibinfo {author} {\bibfnamefont {A.}~\bibnamefont
  {Burov}}, \bibinfo {author} {\bibfnamefont {S.}~\bibnamefont {Nagaitsev}}, \
  and\ \bibinfo {author} {\bibfnamefont {Y.}~\bibnamefont {Derbenev}},\ }\href
  {\doibase 10.1103/PhysRevE.66.016503} {\bibfield  {journal} {\bibinfo
  {journal} {Phys. Rev. E}\ }\textbf {\bibinfo {volume} {66}},\ \bibinfo
  {pages} {016503} (\bibinfo {year} {2002})}\BibitemShut {NoStop}%
\bibitem [{\citenamefont {Kim}(2003)}]{Kim2003}%
  \BibitemOpen
  \bibfield  {author} {\bibinfo {author} {\bibfnamefont {K.-J.}\ \bibnamefont
  {Kim}},\ }\href {\doibase 10.1103/PhysRevSTAB.6.104002} {\bibfield  {journal}
  {\bibinfo  {journal} {Phys. Rev. ST Accel. Beams}\ }\textbf {\bibinfo
  {volume} {6}},\ \bibinfo {pages} {104002} (\bibinfo {year}
  {2003})}\BibitemShut {NoStop}%
\bibitem [{\citenamefont {Bassetti}\ and\ \citenamefont
  {Erskine}(1980)}]{Bassetti1980}%
  \BibitemOpen
  \bibfield  {author} {\bibinfo {author} {\bibfnamefont {M.}~\bibnamefont
  {Bassetti}}\ and\ \bibinfo {author} {\bibfnamefont {G.~A.}\ \bibnamefont
  {Erskine}},\ }\href {https://cds.cern.ch/record/122227} {\emph {\bibinfo
  {title} {Closed expression for the electrical field of a two-dimensional
  Gaussian charge}}},\ \bibinfo {type} {Tech. Rep.}\ \bibinfo {number}
  {CERN-ISR-TH-80-06, ISR-TH-80-06}\ (\bibinfo  {institution} {CERN},\ \bibinfo
  {address} {Geneva},\ \bibinfo {year} {1980})\BibitemShut {NoStop}%
\bibitem [{\citenamefont {Ziemann}(1991)}]{Ziemann1991}%
  \BibitemOpen
  \bibfield  {author} {\bibinfo {author} {\bibfnamefont {V.}~\bibnamefont
  {Ziemann}},\ }in\ \href
  {https://www.slac.stanford.edu/pubs/slacpubs/5500/slac-pub-5582.pdf} {\emph
  {\bibinfo {booktitle} {{7th ICFA Beam Dynamics Workshop}}}},\ \bibinfo
  {series and number} {\bibinfo {number} {SLAC-PUB-5582}}\ (\bibinfo {year}
  {1991})\BibitemShut {NoStop}%
\bibitem [{\citenamefont {Hirata}\ \emph {et~al.}(1993)\citenamefont {Hirata},
  \citenamefont {Moshammer},\ and\ \citenamefont {Ruggiero}}]{Hirata1993}%
  \BibitemOpen
  \bibfield  {author} {\bibinfo {author} {\bibfnamefont {K.}~\bibnamefont
  {Hirata}}, \bibinfo {author} {\bibfnamefont {H.}~\bibnamefont {Moshammer}}, \
  and\ \bibinfo {author} {\bibfnamefont {F.}~\bibnamefont {Ruggiero}},\ }\href
  {https://www-public.slac.stanford.edu/sciDoc/docMeta.aspx?slacPubNumber=SLAC-PUB-10055}
  {\bibfield  {journal} {\bibinfo  {journal} {Part. Accel.}\ }\textbf {\bibinfo
  {volume} {40}},\ \bibinfo {pages} {205} (\bibinfo {year} {1993})}\BibitemShut
  {NoStop}%
\bibitem [{\citenamefont {Hirata}(1995)}]{Hirata1995}%
  \BibitemOpen
  \bibfield  {author} {\bibinfo {author} {\bibfnamefont {K.}~\bibnamefont
  {Hirata}},\ }\href
  {https://journals.aps.org/prl/abstract/10.1103/PhysRevLett.74.2228}
  {\bibfield  {journal} {\bibinfo  {journal} {Phys. Rev. Lett.}\ }\textbf
  {\bibinfo {volume} {74}},\ \bibinfo {pages} {2228} (\bibinfo {year}
  {1995})}\BibitemShut {NoStop}%
\bibitem [{\citenamefont {Papaphilippou}\ and\ \citenamefont
  {Zimmermann}(1999)}]{Papaphilippou1999}%
  \BibitemOpen
  \bibfield  {author} {\bibinfo {author} {\bibfnamefont {Y.}~\bibnamefont
  {Papaphilippou}}\ and\ \bibinfo {author} {\bibfnamefont {F.}~\bibnamefont
  {Zimmermann}},\ }\href
  {https://journals.aps.org/prab/abstract/10.1103/PhysRevSTAB.2.104001}
  {\bibfield  {journal} {\bibinfo  {journal} {Phys. Rev. Special Topics –
  Accel. Beams}\ }\textbf {\bibinfo {volume} {2}},\ \bibinfo {pages} {104001}
  (\bibinfo {year} {1999})}\BibitemShut {NoStop}%
\bibitem [{\citenamefont {Leunissen}\ \emph {et~al.}(2000)\citenamefont
  {Leunissen}, \citenamefont {Schmidt},\ and\ \citenamefont
  {Ripken}}]{Leunissen2000}%
  \BibitemOpen
  \bibfield  {author} {\bibinfo {author} {\bibfnamefont {L.~H.~A.}\
  \bibnamefont {Leunissen}}, \bibinfo {author} {\bibfnamefont {F.}~\bibnamefont
  {Schmidt}}, \ and\ \bibinfo {author} {\bibfnamefont {G.}~\bibnamefont
  {Ripken}},\ }\href
  {https://journals.aps.org/prab/abstract/10.1103/PhysRevSTAB.3.124002}
  {\bibfield  {journal} {\bibinfo  {journal} {Phys. Rev. Special Topics –
  Accel. Beams}\ }\textbf {\bibinfo {volume} {3}},\ \bibinfo {pages} {124002}
  (\bibinfo {year} {2000})}\BibitemShut {NoStop}%
\bibitem [{\citenamefont {Qiang}\ \emph {et~al.}(2004)\citenamefont {Qiang},
  \citenamefont {Furman},\ and\ \citenamefont {Ryne}}]{Qiang2004}%
  \BibitemOpen
  \bibfield  {author} {\bibinfo {author} {\bibfnamefont {J.}~\bibnamefont
  {Qiang}}, \bibinfo {author} {\bibfnamefont {M.~A.}\ \bibnamefont {Furman}}, \
  and\ \bibinfo {author} {\bibfnamefont {R.~D.}\ \bibnamefont {Ryne}},\ }\href
  {https://www.sciencedirect.com/science/article/pii/S0021999104000282}
  {\bibfield  {journal} {\bibinfo  {journal} {J. Comp. Phys.}\ }\textbf
  {\bibinfo {volume} {198}},\ \bibinfo {pages} {278} (\bibinfo {year}
  {2004})}\BibitemShut {NoStop}%
\bibitem [{\citenamefont {Herr}(2006)}]{Brandt2006b}%
  \BibitemOpen
  \bibfield  {author} {\bibinfo {author} {\bibfnamefont {W.}~\bibnamefont
  {Herr}},\ }in\ \href {\doibase 10.5170/CERN-2006-002} {\emph {\bibinfo
  {booktitle} {{Proc., CERN Accelerator School}}}},\ \bibinfo {editor} {edited
  by\ \bibinfo {editor} {\bibfnamefont {D.}~\bibnamefont {Brandt}}}\ (\bibinfo
  {year} {2006})\ pp.\ \bibinfo {pages} {379--410}\BibitemShut {NoStop}%
\bibitem [{\citenamefont {Wienands}\ \emph {et~al.}(2007)\citenamefont
  {Wienands}, \citenamefont {Cai}, \citenamefont {Ecklund}, \citenamefont
  {Seeman},\ and\ \citenamefont {Sullivan}}]{Wienands2007}%
  \BibitemOpen
  \bibfield  {author} {\bibinfo {author} {\bibfnamefont {U.}~\bibnamefont
  {Wienands}}, \bibinfo {author} {\bibfnamefont {Y.}~\bibnamefont {Cai}},
  \bibinfo {author} {\bibfnamefont {S.~D.}\ \bibnamefont {Ecklund}}, \bibinfo
  {author} {\bibfnamefont {J.~T.}\ \bibnamefont {Seeman}}, \ and\ \bibinfo
  {author} {\bibfnamefont {M.~K.}\ \bibnamefont {Sullivan}},\ }in\ \href
  {\doibase 10.1109/PAC.2007.4440330} {\emph {\bibinfo {booktitle} {2007 IEEE
  Part. Accel. Conf. (PAC)}}}\ (\bibinfo {year} {2007})\ pp.\ \bibinfo {pages}
  {37--41}\BibitemShut {NoStop}%
\bibitem [{\citenamefont {Olive}\ \emph
  {et~al.}(date{\natexlab{a}})\citenamefont {Olive} \emph
  {et~al.}}]{Olive2014a}%
  \BibitemOpen
  \bibfield  {author} {\bibinfo {author} {\bibfnamefont {K.~A.}\ \bibnamefont
  {Olive}} \emph {et~al.},\ }\href
  {https://pdg.lbl.gov/2015/reviews/rpp2015-rev-hep-collider-params.pdf}
  {\bibfield  {journal} {\bibinfo  {journal} {30. High-energy collider
  parameters, Chin. Phys. C}\ }\textbf {\bibinfo {volume} {38}},\ \bibinfo
  {pages} {090001} (\bibinfo {year} {2014 and 2015
  update}{\natexlab{a}})}\BibitemShut {NoStop}%
\bibitem [{\citenamefont {Accardi}\ \emph {et~al.}(2016)\citenamefont {Accardi}
  \emph {et~al.}}]{Accardi2016}%
  \BibitemOpen
  \bibfield  {author} {\bibinfo {author} {\bibfnamefont {A.}~\bibnamefont
  {Accardi}} \emph {et~al.},\ }\href
  {https://link.springer.com/article/10.1140/epja/i2016-16268-9} {\bibfield
  {journal} {\bibinfo  {journal} {Eur. 554 Phys. J. A}\ }\textbf {\bibinfo
  {volume} {52}},\ \bibinfo {pages} {268} (\bibinfo {year} {2016})}\BibitemShut
  {NoStop}%
\bibitem [{\citenamefont {Adam}\ \emph {et~al.}(2021)\citenamefont {Adam} \emph
  {et~al.}}]{Adam2021}%
  \BibitemOpen
  \bibfield  {author} {\bibinfo {author} {\bibfnamefont {J.}~\bibnamefont
  {Adam}} \emph {et~al.},\ }\href
  {https://www.bnl.gov/ec/files/EIC_CDR_Final.pdf} {\enquote {\bibinfo {title}
  {Electron-ion collider at brookhaven national laboratory, conceptual design
  report},}\ } (\bibinfo {year} {2021})\BibitemShut {NoStop}%
\bibitem [{\citenamefont {Zimmermann}(2018)}]{Zimmermann2018}%
  \BibitemOpen
  \bibfield  {author} {\bibinfo {author} {\bibfnamefont {F.}~\bibnamefont
  {Zimmermann}},\ }\href {\doibase 10.1016/j.nima.2018.01.034} {\bibfield
  {journal} {\bibinfo  {journal} {Nucl. Instrum. Methods Phys. Res., Sect. A}\
  }\textbf {\bibinfo {volume} {909}},\ \bibinfo {pages} {33} (\bibinfo {year}
  {2018})}\BibitemShut {NoStop}%
\bibitem [{\citenamefont {Herr}\ and\ \citenamefont
  {Muratori}(2006)}]{Brandt2006a}%
  \BibitemOpen
  \bibfield  {author} {\bibinfo {author} {\bibfnamefont {W.}~\bibnamefont
  {Herr}}\ and\ \bibinfo {author} {\bibfnamefont {B.}~\bibnamefont
  {Muratori}},\ }in\ \href {\doibase 10.5170/CERN-2006-002} {\emph {\bibinfo
  {booktitle} {{Proc., CERN Accelerator School}}}},\ \bibinfo {editor} {edited
  by\ \bibinfo {editor} {\bibfnamefont {D.}~\bibnamefont {Brandt}}}\ (\bibinfo
  {year} {2006})\ pp.\ \bibinfo {pages} {361--378}\BibitemShut {NoStop}%
\bibitem [{\citenamefont {Olive}\ \emph
  {et~al.}(date{\natexlab{b}})\citenamefont {Olive} \emph
  {et~al.}}]{Olive2014b}%
  \BibitemOpen
  \bibfield  {author} {\bibinfo {author} {\bibfnamefont {K.~A.}\ \bibnamefont
  {Olive}} \emph {et~al.},\ }\href
  {https://pdg.lbl.gov/2015/reviews/rpp2015-rev-accel-phys-colliders.pdf}
  {\bibfield  {journal} {\bibinfo  {journal} {29. Accelerator physics of
  colliders, 29.1. Luminosity, Chin. Phys. C}\ }\textbf {\bibinfo {volume}
  {38}},\ \bibinfo {pages} {090001} (\bibinfo {year} {2014 and 2015
  update}{\natexlab{b}})}\BibitemShut {NoStop}%
\bibitem [{\citenamefont {Piot}\ \emph {et~al.}(2006)\citenamefont {Piot},
  \citenamefont {Sun},\ and\ \citenamefont {Kim}}]{Piot2006}%
  \BibitemOpen
  \bibfield  {author} {\bibinfo {author} {\bibfnamefont {P.}~\bibnamefont
  {Piot}}, \bibinfo {author} {\bibfnamefont {Y.-E.}\ \bibnamefont {Sun}}, \
  and\ \bibinfo {author} {\bibfnamefont {K.-J.}\ \bibnamefont {Kim}},\ }\href
  {\doibase 10.1103/PhysRevSTAB.9.031001} {\bibfield  {journal} {\bibinfo
  {journal} {Phys. Rev. ST Accel. Beams}\ }\textbf {\bibinfo {volume} {9}},\
  \bibinfo {pages} {031001} (\bibinfo {year} {2006})}\BibitemShut {NoStop}%
\bibitem [{\citenamefont {Borland}(2006)}]{Borland2006}%
  \BibitemOpen
  \bibfield  {author} {\bibinfo {author} {\bibfnamefont {M.}~\bibnamefont
  {Borland}},\ }\href {\doibase 10.1016/j.nima.2005.10.076} {\bibfield
  {journal} {\bibinfo  {journal} {Nucl. Instrum. Meth. A}\ }\textbf {\bibinfo
  {volume} {557}},\ \bibinfo {pages} {230} (\bibinfo {year}
  {2006})}\BibitemShut {NoStop}%
\bibitem [{\citenamefont {Tsumaki}\ and\ \citenamefont
  {Kumagai}(2006)}]{Tsumaki2006}%
  \BibitemOpen
  \bibfield  {author} {\bibinfo {author} {\bibfnamefont {K.}~\bibnamefont
  {Tsumaki}}\ and\ \bibinfo {author} {\bibfnamefont {N.}~\bibnamefont
  {Kumagai}},\ }\href {\doibase 10.1016/j.nima.2006.06.030} {\bibfield
  {journal} {\bibinfo  {journal} {Nucl. Instrum. Meth. A}\ }\textbf {\bibinfo
  {volume} {565}},\ \bibinfo {pages} {394} (\bibinfo {year}
  {2006})}\BibitemShut {NoStop}%
\bibitem [{\citenamefont {Hao}\ and\ \citenamefont {Ptitsyn}(2010)}]{Hao2010}%
  \BibitemOpen
  \bibfield  {author} {\bibinfo {author} {\bibfnamefont {Y.}~\bibnamefont
  {Hao}}\ and\ \bibinfo {author} {\bibfnamefont {V.}~\bibnamefont {Ptitsyn}},\
  }\href {\doibase 10.1103/PhysRevSTAB.13.071003} {\bibfield  {journal}
  {\bibinfo  {journal} {Phys. Rev. ST Accel. Beams}\ }\textbf {\bibinfo
  {volume} {13}},\ \bibinfo {pages} {071003} (\bibinfo {year}
  {2010})}\BibitemShut {NoStop}%
\bibitem [{\citenamefont {Hao}\ \emph {et~al.}(2013)\citenamefont {Hao},
  \citenamefont {Litvinenko},\ and\ \citenamefont {Ptitsyn}}]{Hao2013}%
  \BibitemOpen
  \bibfield  {author} {\bibinfo {author} {\bibfnamefont {Y.}~\bibnamefont
  {Hao}}, \bibinfo {author} {\bibfnamefont {V.~N.}\ \bibnamefont {Litvinenko}},
  \ and\ \bibinfo {author} {\bibfnamefont {V.}~\bibnamefont {Ptitsyn}},\ }\href
  {\doibase 10.1103/PhysRevSTAB.16.101001} {\bibfield  {journal} {\bibinfo
  {journal} {Phys. Rev. ST Accel. Beams}\ }\textbf {\bibinfo {volume} {16}},\
  \bibinfo {pages} {101001} (\bibinfo {year} {2013})}\BibitemShut {NoStop}%
\bibitem [{\citenamefont {Heifets}\ \emph {et~al.}(1990)\citenamefont
  {Heifets}, \citenamefont {Krafft},\ and\ \citenamefont
  {Fripp}}]{Heifets1990}%
  \BibitemOpen
  \bibfield  {author} {\bibinfo {author} {\bibfnamefont {S.~A.}\ \bibnamefont
  {Heifets}}, \bibinfo {author} {\bibfnamefont {G.~A.}\ \bibnamefont {Krafft}},
  \ and\ \bibinfo {author} {\bibfnamefont {M.}~\bibnamefont {Fripp}},\ }\href
  {\doibase https://doi.org/10.1016/0168-9002(90)90704-A} {\bibfield  {journal}
  {\bibinfo  {journal} {Nucl. Instrum. Methods Phys. Res., Sect. A}\ }\textbf
  {\bibinfo {volume} {295}},\ \bibinfo {pages} {286} (\bibinfo {year}
  {1990})}\BibitemShut {NoStop}%
\bibitem [{\citenamefont {Li}\ \emph {et~al.}(2001)\citenamefont {Li},
  \citenamefont {Yunn}, \citenamefont {Lebedev},\ and\ \citenamefont
  {Bisognano}}]{Li2001}%
  \BibitemOpen
  \bibfield  {author} {\bibinfo {author} {\bibfnamefont {R.}~\bibnamefont
  {Li}}, \bibinfo {author} {\bibfnamefont {B.~C.}\ \bibnamefont {Yunn}},
  \bibinfo {author} {\bibfnamefont {V.}~\bibnamefont {Lebedev}}, \ and\
  \bibinfo {author} {\bibfnamefont {J.~J.}\ \bibnamefont {Bisognano}},\ }in\
  \href {\doibase 10.1109/PAC.2001.987260} {\emph {\bibinfo {booktitle}
  {PACS2001. Proc. 2001 Part. Accel. Conf. (Cat. No.01CH37268)}}},\
  Vol.~\bibinfo {volume} {3}\ (\bibinfo  {publisher} {IEEE},\ \bibinfo {year}
  {2001})\ pp.\ \bibinfo {pages} {2014--2016}\BibitemShut {NoStop}%
\bibitem [{\citenamefont {Fayet}(2007)}]{Fayet2007}%
  \BibitemOpen
  \bibfield  {author} {\bibinfo {author} {\bibfnamefont {P.}~\bibnamefont
  {Fayet}},\ }\href {\doibase 10.1103/PhysRevD.75.115017} {\bibfield  {journal}
  {\bibinfo  {journal} {Phys. Rev. D}\ }\textbf {\bibinfo {volume} {75}},\
  \bibinfo {pages} {115017} (\bibinfo {year} {2007})},\ \Eprint
  {http://arxiv.org/abs/hep-ph/0702176} {arXiv:hep-ph/0702176} \BibitemShut
  {NoStop}%
\bibitem [{\citenamefont {Raimondi}\ and\ \citenamefont
  {Seryi}(2001)}]{Raimondi2001}%
  \BibitemOpen
  \bibfield  {author} {\bibinfo {author} {\bibfnamefont {P.}~\bibnamefont
  {Raimondi}}\ and\ \bibinfo {author} {\bibfnamefont {A.}~\bibnamefont
  {Seryi}},\ }\href {\doibase 10.1103/PhysRevLett.86.3779} {\bibfield
  {journal} {\bibinfo  {journal} {Phys. Rev. Lett.}\ }\textbf {\bibinfo
  {volume} {86}},\ \bibinfo {pages} {3779} (\bibinfo {year}
  {2001})}\BibitemShut {NoStop}%
\bibitem [{\citenamefont {Morozov}\ \emph {et~al.}(2013)\citenamefont
  {Morozov}, \citenamefont {Derbenev}, \citenamefont {Lin},\ and\ \citenamefont
  {Johnson}}]{Morozov2013}%
  \BibitemOpen
  \bibfield  {author} {\bibinfo {author} {\bibfnamefont {V.~S.}\ \bibnamefont
  {Morozov}}, \bibinfo {author} {\bibfnamefont {Y.~S.}\ \bibnamefont
  {Derbenev}}, \bibinfo {author} {\bibfnamefont {F.}~\bibnamefont {Lin}}, \
  and\ \bibinfo {author} {\bibfnamefont {R.~P.}\ \bibnamefont {Johnson}},\
  }\href {\doibase 10.1103/PhysRevSTAB.16.011004} {\bibfield  {journal}
  {\bibinfo  {journal} {Phys. Rev. ST Accel. Beams}\ }\textbf {\bibinfo
  {volume} {16}},\ \bibinfo {pages} {011004} (\bibinfo {year}
  {2013})}\BibitemShut {NoStop}%
\end{thebibliography}%

\end{document}